\documentclass[aps,twocolumn,unsortedaddress,a4paper,longbibliography]{revtex4-1}
\usepackage{amsmath}
\usepackage{amsfonts}
\usepackage{amssymb}
\usepackage{graphicx}
\usepackage{epstopdf}
\usepackage{multirow}
\usepackage{tabularx}
\usepackage{ragged2e}

\begin{document}

\title{Avoidance and Coalescence of Delamination Patterns}

\author{Zoe Budrikis}
\affiliation{ISI Foundation, Via Alassio 11/c, 10126 Torino, Italy}
\author{Alessandro L. Sellerio}
\affiliation{CNR-IENI, Via R. Cozzi 53, 20125 Milano, Italy}
\author{Zsolt Bertalan}
\affiliation{ISI Foundation, Via Alassio 11/c, 10126 Torino, Italy}
\author{Stefano Zapperi}
\affiliation{CNR-IENI, Via R. Cozzi 53, 20125 Milano, Italy}
\affiliation{ISI Foundation, Via Alassio 11/c, 10126 Torino, Italy}

\begin{abstract}
Delamination of coatings and thin films from substrates generates a fascinating variety of patterns, from
circular blisters to wrinkles and labyrinth domains, in a way that is not completely understood. We
report on large-scale numerical simulations of the universal problem of avoidance and coalescence
of delamination wrinkles, focusing on a case study of graphene sheets on patterned substrates.
By nucleating and growing wrinkles in a controlled way, we are able to characterize how their interactions,
mediated by long-range stress fields, determine their formation and morphology. We
also study how the interplay between geometry and stresses drives a universal transition from conformation to delamination
when sheets are deposited on particle-decorated substrates. Our results are directly applicable to strain
engineering of graphene and also uncover universal phenomena observed at all scales, as for example
in geomembrane deposition.
\end{abstract}

\maketitle

\section{Introduction}
Delamination of a thin sheet from a substrate may be an undesirable effect, for example when plastic coatings peel from glass, but controlled delamination is of interest for a range of applications from flexible electronics~\cite{Sun2006} to building micro- and nanofluidic devices~\cite{Malachias2008, Mei2010}.
It is therefore of great technological importance to understand the role played by sheet properties and substrate geometry in determining delamination and sheet morphology.
In addition, delamination blisters and wrinkles form a rich variety of complex patterns, which are the subject of ongoing research. Examples include circular blisters, ``telephone cord'' buckles, parallel and radially organized wrinkles, and labyrinth patterns \cite{Audoly1999,Coupeau2002, Kwon2005, Edmondson2006, Jagla2007, Vandeparre2008, Aoyanagi2010, Annabattula2011, Faou2012, Yu2013}. In the case of graphene sheets,  patterning into networks of wrinkles using nanopillars or particles on the substrate has been experimentally demonstrated~\cite{Tomori2011, NeekAmal2012, Yamamoto2012, Zhu2014}. Previous theoretical work on thin sheet delamination has focused on the properties of a single blister or wrinkle, looking for example at how blister growth is affected by substrate interactions and elastic instabilities~\cite{Audoly1999, Aoyanagi2010, ViolaKusminskiy2011, Faou2012}. On the other hand, interactions between delaminated regions are fundamental as strikingly illustrated by recent experiments by Yamamoto {\it et al}~\cite{Yamamoto2012} on graphene sheets.

Graphene wrinkles are seen to form pairs that approach each other but do not coalesce~\cite{Yamamoto2012} (see Fig.~\ref{yamamoto_comparison}), as also confirmed by atomistic simulations \cite{Zhu2014}. Avoiding wrinkle pairs are not just a curiosity of graphene sheets on patterned substrates, and in fact can be found not only in other thin films but even in macroscopic sheets such as geomembranes used to line landfills. Examples from the physics and engineering experimental literature are listed in Table~\ref{wrinkles_in_literature}. Despite their ubiquitous nature, the physics leading to avoiding wrinkle pairs is largely unexplored. It is remarkable that similar universal avoiding patterns are widely observed in cracks. ``En passant" cracks in films and plates \cite{Kranz1979} have been recently shown to be display universal morphology, independent on the material properties  \cite{Fender2010}. Do similar universal laws apply to avoiding wrinkle pairs?

Here we report a large-scale simulation study of avoiding and coalescing delamination wrinkles in graphene. The phenomena we describe are universal, but we have chosen to focus on graphene as a case study because it is an exemplary thin sheet, being truly two-dimensional and having a  remarkably high ratio of bending rigidity ($\sim 1$ eV) to tensile rigidity ($\sim 10^3$ eV/nm$^2$)~\cite{Fasolino2007, Lee2012, Hartmann2013}, which allows narrow buckled morphological features to form. 
This makes it an ideal model system in which to explore the mechanics of membranes and fascinating in its own right~\cite{Novoselov2004,CastroNeto2009,Novoselov2012}.
Furthermore, strain engineering of graphene promises control of electronic, optical and thermal properties, making it useful for a range of applications~\cite{Pereira2009, Wei2011,Ni2014}, so that understanding the morphology of graphene sheets on substrates is of value for building devices. 

We introduce a coarse grained model for mono- and multilayer graphene, which we validate by comparing simulations of micron square sheets with existing experimental results~\cite{Yamamoto2012}. We then report on simulations of controlled growth of interacting delamination wrinkles, which are inspired by Fender {\it et al}'s experiments on cracks in plates~\cite{Fender2010}. These reveal the role played by long range interactions in determining the shapes of wrinkles, and whether they form avoiding pairs or coalesce. We also discuss how coalescence of wrinkles can be used to induce delamination in rectangular regions. Our simulation studies illustrate the importance of free edges on the sheet for allowing lateral motion, without which it is impossible to model experimentally-observed phenomena correctly.

\begin{table*}[htp]
  \begin{tabularx}{\linewidth}{ l >{\RaggedRight}X >{\RaggedRight}X >{\RaggedRight}X >{\RaggedRight}X l }
    \hline
	Reference & Figure &   Material & Substrate & Thickness\; & Wrinkle width \\
	\hline
	\cite{Yamamoto2012} & Fig.~1d,e. \footnote{c.f. Fig.~1 of main text.} & Graphene & Silicon/silica nano\-particles & Single atom & $\sim 7.4$ nm \footnote{Based on the average wrinkle width being approximately the nanoparticle diameter \cite{Yamamoto2012}.}\\
	\cite{Sutter2013} & Fig.~2b, upper centre. & Graphene (exposed to O$_2$) & Ru(1000) & Single atom & $\sim 1$ nm\\
	\cite{Malachias2008} & Fig.~2b, lower right panel. & In$_{0.2}$Ga$_{0.8}$As & AlAs with pits & $10$ nm & $\sim 1$ $\mu$m\\
	\cite{Matuda1981} & Fig.~2b, e.g, near upper edge of scale bar. & C & Glass & $80$ nm & $\sim 5$ $\mu$m\\
	\cite{Kwon2005} & Fig.~1b. & ZnO & Silicon & $85$ nm & $\sim 0.5$ $\mu$m\\
	\cite{Coupeau2002} & Fig.~1a. & Ni & Polycarbonate &  $350$ nm  & $\sim 10$ $\mu$m \\
	\cite{Take2007} & Fig.~5, near image centre. & High-density polyethylene & Clay & $1.5$ mm & $0.31\pm0.06$ m \footnote{As quoted in Ref.~\cite{Take2007}.} \\
	\hline
  \end{tabularx}
  \caption{ \label{wrinkles_in_literature} Examples of avoiding pairs of wrinkles in the literature, sorted by sheet thickness. This is not intended to be an exhaustive list, but rather to give a sense of the variety of situations in which avoiding pairs occur.}
\end{table*}

\section{Coarse grained model of an $n$-layer graphene sheet}
While atomistic simulations provide a useful benchmark, their computational intensity limits their usefulness for studies of mesoscopic systems. Furthermore, we are primarily interested in universal phenomena, rather than graphene-specific behavior. We therefore develop a linearized model that captures the essential physics of an ultrathin sheet interacting with a patterned substrate, and show below that it is capable of describing the mechanical behavior of graphene in which we are interested. 
 
We model the graphene sheet as a triangular lattice of particles connected by springs with rest length $a=5$\AA, which provide a stretching energy
\begin{equation}
E=k_s (x-a)^2
\end{equation} 
where $x$ is the spring length and $k_s=n k_s^*$ is the stiffness for a graphene sheet composed by $n$ layers, with  
$k_s^*=20$ eV/\AA$^2$ \cite{Lee2012, Hartmann2013}. 
Bending rigidity is implemented as an energy that depends on  the angle $\phi$ between two neighbouring triangles and is minimized when the sheet is flat,
\begin{equation}
E=k_b (\phi - \pi)^2
\end{equation} 
where $k_b= k_b^* n^3$ is the bending stiffness with $k_b^*=1$ eV/rad$^2$ (Ref~\cite{Fasolino2007,  Poot2008}). 
A distinct advantage of these simulations over atomistic simulations is that sheet thickness enters as a change in parameters, without a change in the number of particles simulated. This allows us to study any number of layers as long as the assumptions of thin-sheet elasticity hold.

The sheet is attracted to the substrate {\it via} a downward force applied on all particles that is independent of height, but the substrate itself has a radial repulsive interaction with the sheet, to form a hard barrier. The strength of these forces has been selected to give qualitative agreement with experiments (see below). This substrate interaction guarantees a substrate solid enough to deposit particles on, and behaves in a reasonable way even if the specific choice is rather arbitrary

Open boundary conditions are used, so that the sheet is able to expand and shrink laterally. As discussed below, an adequate accounting for lateral deformation of the sheet is essential to capture the physics of the system. In order to reduce spurious edge effects and allow for effects of  friction between the sheet and substrate, particles near the sheet edge are coupled to their original lateral positions with harmonic springs, to constrain their lateral (but not vertical) motion. The simulation is performed using the molecular dynamics package LAMMPS~\cite{Plimpton1995} and is conducted at zero temperature, with damping applied to slowly drain energy from the system. Further details are given in Appendix~\ref{lammps}.

\section{Results}

\subsection{Wrinkle networks on decorated substrates}
\begin{figure}[htp]
\includegraphics[width=\columnwidth]{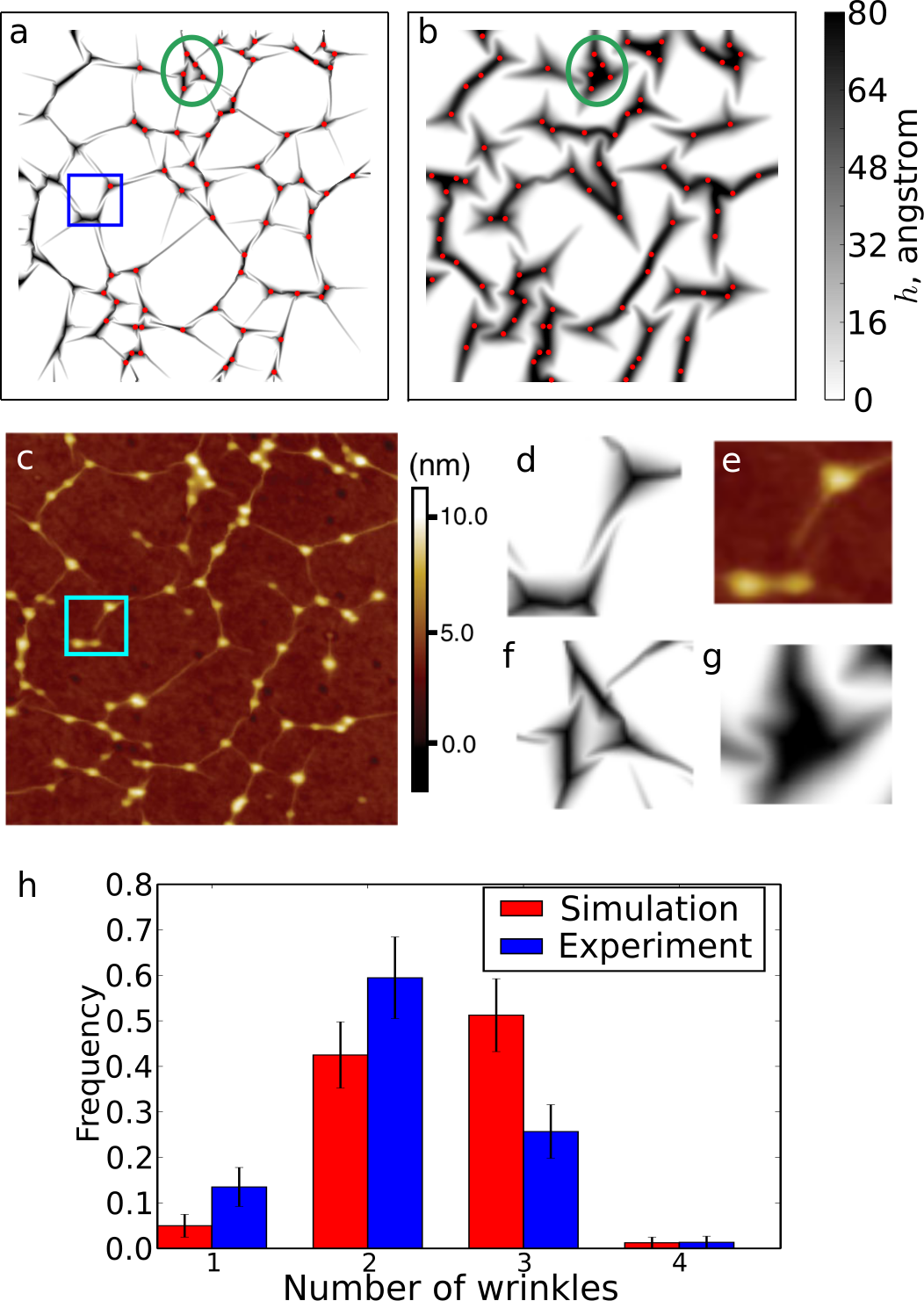} 
\caption{\label{yamamoto_comparison}
Simulated $1\times 1$ $\mu$m$^2$ (a) monolayer and (b) 5-layer graphene sheet on a substrate decorated with spherical particles of diameter $8$ nm, placed in positions determined by digitization of the experimental micrograph shown in panel (c). Red dots indicate the particle centres.
(c) Atomic force micrograph from Ref.~\cite{Yamamoto2012} (DOI:10.1103/PhysRevX.2.041018), 
showing a $1\times 1$ $\mu$m$^2$ region of graphene exfoliated onto a substrate decorated with nanoparticles of mean diameter $7.4$nm. 
(d,e) Close-ups of an example avoiding pair of wrinkles found in experiment and simulation, as highlighted by the blue boxes in panels (a) and (c).
(f,g) Close-ups of a region in which the monolayer sheet conforms to the substrate but the 5-layer sheet delaminates, as indicated by the green ellipses in panels (a) and (b).
(h) Histograms of the number of wrinkles emanating from each particle, as counted from the images in panels (a) and (c). Error bars correspond to the square root of each count.}
\end{figure}

We verify our model through comparison with experimental studies of graphene sheets deposited on substrates decorated with nanoparticles of diameter $\sim 10$ nm, as published recently by Yamamoto {\it et al}~\cite{Yamamoto2012}. 
We simulate $1 \times 1$ $\mu$m$^2$ regions of monolayer and 5-layer graphene, with particles of radius $8$~nm placed at positions determined by digitizing the atomic force micrograph that was obtained by Yamamoto {\it et al}~\cite{Yamamoto2012} for a system with nanoparticle density of 90 $\mu$m$^{-2}$. As seen in Fig.~\ref{yamamoto_comparison}, the agreement between monolayer simulation and experiment is good, with both systems displaying similar networks of long narrow wrinkles connecting particles. We note that agreement is only expected in the center of the region simulated, because  the experimental image is a sample of a larger region so that wrinkles near the edges are affected by particles that are not shown in the limited field of view.

Importantly, our linearized simulations exhibit avoiding pairs of wrinkles, an example of which is highlighted in Fig.~\ref{yamamoto_comparison}(d,e). The pairs are stable for long simulation times (up to at least $2\times 10^6$ time steps) and occur under a range of downward forces on the sheet. In addition, like Yamamoto {\it et al}, we find that increasing the thickness of the graphene sheet hinders its conformation to the substrate (compare our Fig.~\ref{yamamoto_comparison}(a,b) with Fig.~6 of Ref.~\cite{Yamamoto2012}). This is especially apparent in regions where close-spaced particles are positioned on the vertices of a polygon, as in the region indicated in Fig.~\ref{yamamoto_comparison}(a,b). As highlighted in Fig.~\ref{yamamoto_comparison}(f,g), the sheet in this region ``pops up'' when its thickness (and hence ratio of bending to stretching costs) increases. In the rest of this paper, we study these two characteristic behaviors in greater detail.

Beyond visual comparisons, we compare simulation and experiment statistically, by measuring the distribution of the number of wrinkles emanating from each particle for the grpahene monolayers, as plotted in Fig.~\ref{yamamoto_comparison}(h).  The agreement is good, although the simulations have a slightly larger mean number of wrinkles per particle ($2.48$ for simulation, {\it vs} $2.15$ for experiment). However, this may simply be because the lower noise levels in the simulation data allow more wrinkles to be detected.

\subsection{Morphology of wrinkle pairs}
\begin{figure}
\includegraphics[width=\columnwidth]{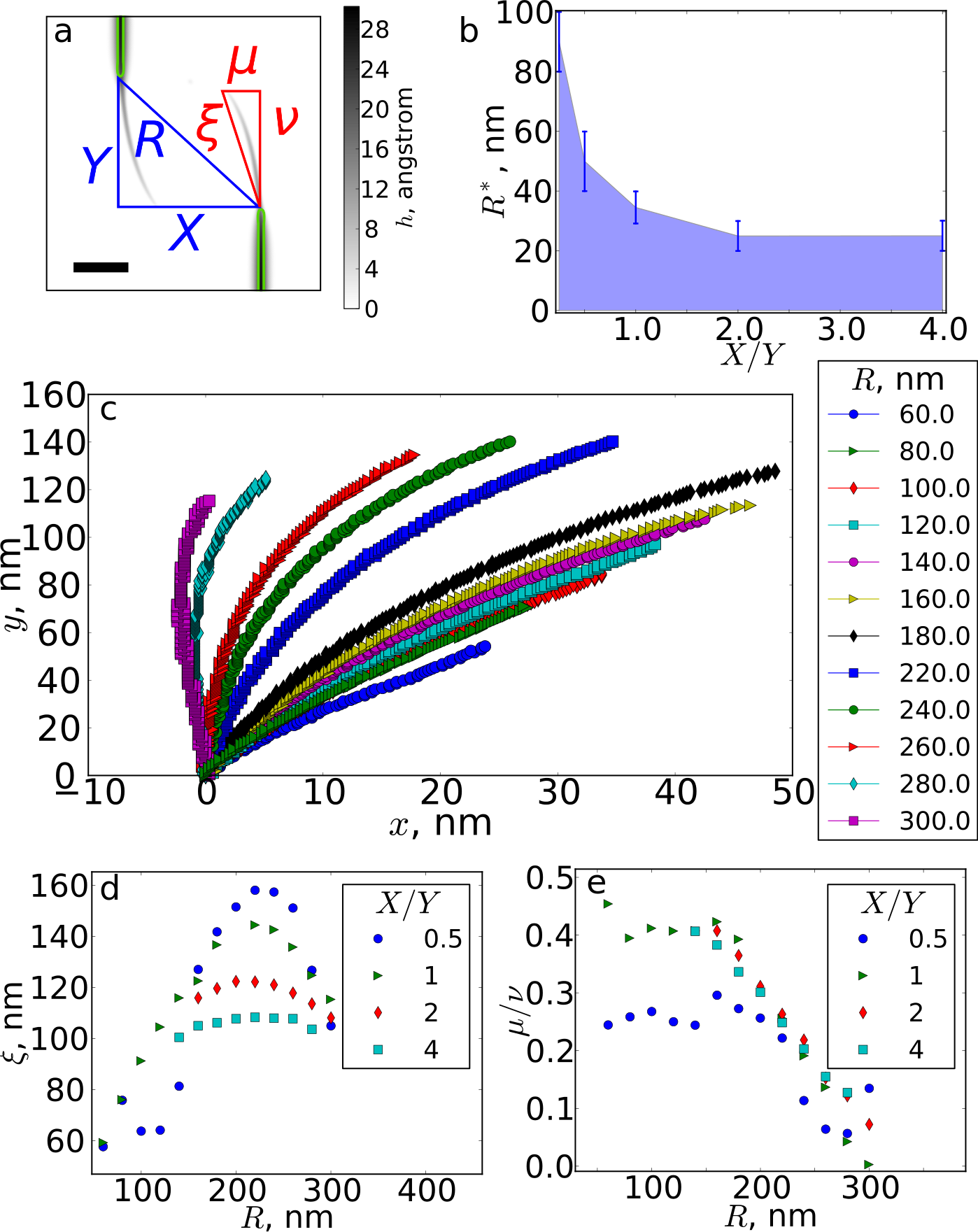} 
\caption{\label{wrinkle_shape} 
(a) Schematic of wrinkle interaction tests shown on a close-up of the central region of a sheet. Elongated ``wrinkle nuclei'', indicated by green lines, are lifted with constant force, to cause wrinkles to propagate towards the centre of the sheet. We investigate how changing the nucleus separations $X$ and $Y$ affects the wrinkle shape, which is characterized by $\mu$ and $\nu$. The scale bar indicates $50$ nm.
(b) Whether two wrinkles join or remain separate depends on nucleus separations, here characterized by aspect ratio $X/Y$ and distance. When the nuclei are nearly head on ($X/Y$ small), even distant pairs join, but the maximum distance for joining drops off rapidly as the aspect ratio is increased.
(c) Wrinkle shapes for $X=Y$, for a range of $R$ values. The wrinkle nucleus is excluded, and wrinkles are transformed so the end of the nucleus (the base of the wrinkle) is situated at $(x=0,y=0)$ and the tip of the wrinkle is to the right of its base.
(d) The extension of a wrinkle, $\xi=\sqrt{\mu^2+\nu^2}$, is maximised for a separation of $\sim 220$ nm, but decreases with increasing aspect ratio.
(e) Remarkably, when $X/Y\geq 1$, the aspect ratio of the wrinkle, $\mu/\nu$ depends only on nucleus tip separation $R$.}
\end{figure}

We first study the growth and interactions of wrinkles, with a focus on avoiding pairs. In order to characterize and understand these pairs, we require a controlled method for creating and growing them.
Inspired by studies of interacting cracks grown from a pair of notches~\cite{Fender2010}, we study propagating and interacting wrinkles using simulations in which an initially flat sheet of dimensions $600\times600$ nm$^2$ is subject to upward pulling in well-defined ``wrinkle nuclei'' regions. The tips of the nuclei are separated by distances $X$ and $Y$ as defined in Fig.~\ref{wrinkle_shape}(a), which are our control parameters. We set the upward force to $0.01$~eV/\AA, a factor of $10$ larger than the downward force that is applied to the whole sheet. Because the nuclei are long and narrow and the sheet has high stretching cost, the upward driving primarily causes the wrinkle to propagate forward, with minimal lateral broadening, as shown in Fig.~\ref{wrinkle_shape}. We run each simulation until a stable configuration is attained, which occurs when the applied upward forces are balanced by stresses in the sheet. We have also tested simulations where the upward force is increased slowly, rather than turned on instantaneously, and find that although this allows smoother initial wrinkle growth, the final configurations attained are not strongly affected, as seen in example movies in Supplemental Material~\cite{SuppMat}. We have also tested changing the nucleus orientation relative to the sheet edges (and therefore lattice), but do not find any lattice effects.

Whereas an isolated wrinkle propagates directly forward, parallel to its nucleus,  nearby wrinkles interact via long-range stresses in the sheet. This leads to wrinkles either joining or forming an avoiding pair. For fixed driving conditions, we find that whether a pair of wrinkles join is controlled by the separation of their nuclei. As shown in Fig.~\ref{wrinkle_shape}(b), this can be characterized by aspect ratio $X/Y$ and distance $R$. When $X>Y$, the joining condition is simply a distance threshold of $\sim 30$ nm. However, for $X<Y$, wrinkle merging becomes more favourable and the minimum distance required for the formation of an avoiding pair increases rapidly.

In addition to the conditions for avoiding pair formation, we also characterize how the morphology of such pairs depends on wrinkle nuclei separation. Typical wrinkle shapes are shown in Fig.~\ref{wrinkle_shape}(c), for fixed $X/Y=1$. For small $R$, wrinkles are directed toward each other, but for larger $R$ they are initially repelled before being attracted.
To quantitatively describe the wrinkle shapes, we focus on two characteristics which are indicated in Fig.~\ref{wrinkle_shape}(a): wrinkle extension, which is the distance $\xi = \sqrt{\mu^2+\nu^2}$ from base to tip, and aspect ratio $\mu/\nu$. For fixed $X/Y$, $\xi$ depends on $R$ and is maximized at a distance $\sim 230$ nm, as shown in Fig.~\ref{wrinkle_shape}(d). As shown in Fig.~\ref{wrinkle_shape}(e), the maximum extension decreases with increasing $X/Y$, so that avoiding pairs of wrinkles formed by nearly head-on nuclei are substantially longer than wrinkles formed from nuclei that are separated in the $X$ direction. 
Remarkably, for nucleus separations $R\gtrsim200$ nm, $X/Y$ has little effect on the aspect ratio $\mu/\nu$ of wrinkles, which decreases approximately linearly with $R$.

\subsection{Interactions of wrinkle pairs}

\begin{figure}
\includegraphics[width=\columnwidth]{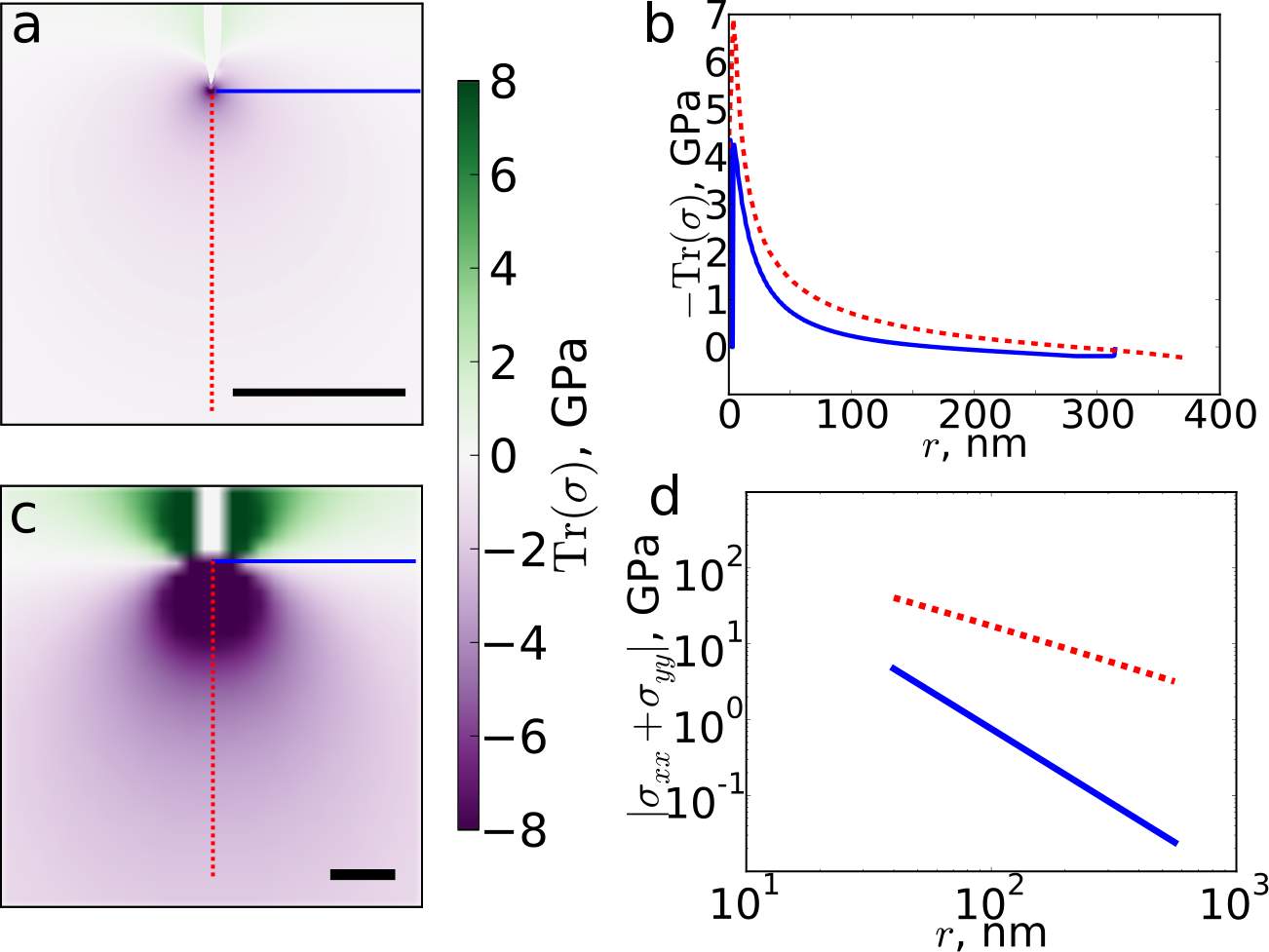} 
\caption{\label{single_wrinkle} 
(a) Trace of the stress tensor from a single wrinkle in the coarse grained model, showing tension behind the tip and compression in front, concentrated at the tip. The scale bar represents $100$ nm. 
(b) Slices of the stress trace parallel and perpendicular to the wrinkle direction, as indicated by red and blue lines in panel (a). 
(c) $\sigma_{xx}+\sigma_{yy}$ for an infinite sheet with an idealized wrinkle imposed, as described in Appendix~\ref{FvK}.
(d) Slices of $\sigma_{xx}+\sigma_{yy}$ parallel and perpendicular to the wrinkle direction, as indicated by red and blue lines in panel (c). 
For clarity, in panels (a) and (c) the interiors of the wrinkles, defined by a height threshold of 2 \AA, are not shown.  }
\end{figure}

\label{sec_single_wrinkle}
In order to understand how wrinkle interactions affect their shape, we characterize the stresses induced by wrinkles, {\it via} the trace of the stress tensor $\mathrm{Tr}(\sigma)=\sigma_{xx}+\sigma_{yy}+\sigma_{zz}$. (The calculation of the stress tensor is described in Appendix~\ref{stress_calc}.) We first consider a single wrinkle. As shown in Fig.~\ref{single_wrinkle}, the stresses emanating from the tip of a wrinkle are compressive ($\mathrm{Tr}(\sigma)<0$), which causes the wrinkle to grow {\it via} buckling of the sheet, so that deformation occurs as bending rather than stretching which is highly unfavourable for graphene. 

As indicated in Fig.~\ref{single_wrinkle}(b), the stress distribution displays an angular dependence and compressive stresses are largest along a line parallel to the wrinkle. We have also calculated the stresses generated by an idealized wrinkle in an infinite sheet under the assumption of zero in-plane deformation by imposing a wrinkle shape and numerically calculating the stresses using standard thin-plate elasticity, as described in Appendix~\ref{FvK}.  As shown in Fig.~\ref{single_wrinkle}(c,d), the idealized wrinkle also generates a stress field with an angular dependence. Far from the wrinkle tip, the stresses decay as a power law with exponent $1$ along the wrinkle axis and $2$ perpendicular to the wrinkle axis.

\begin{figure}
\includegraphics[width=\columnwidth]{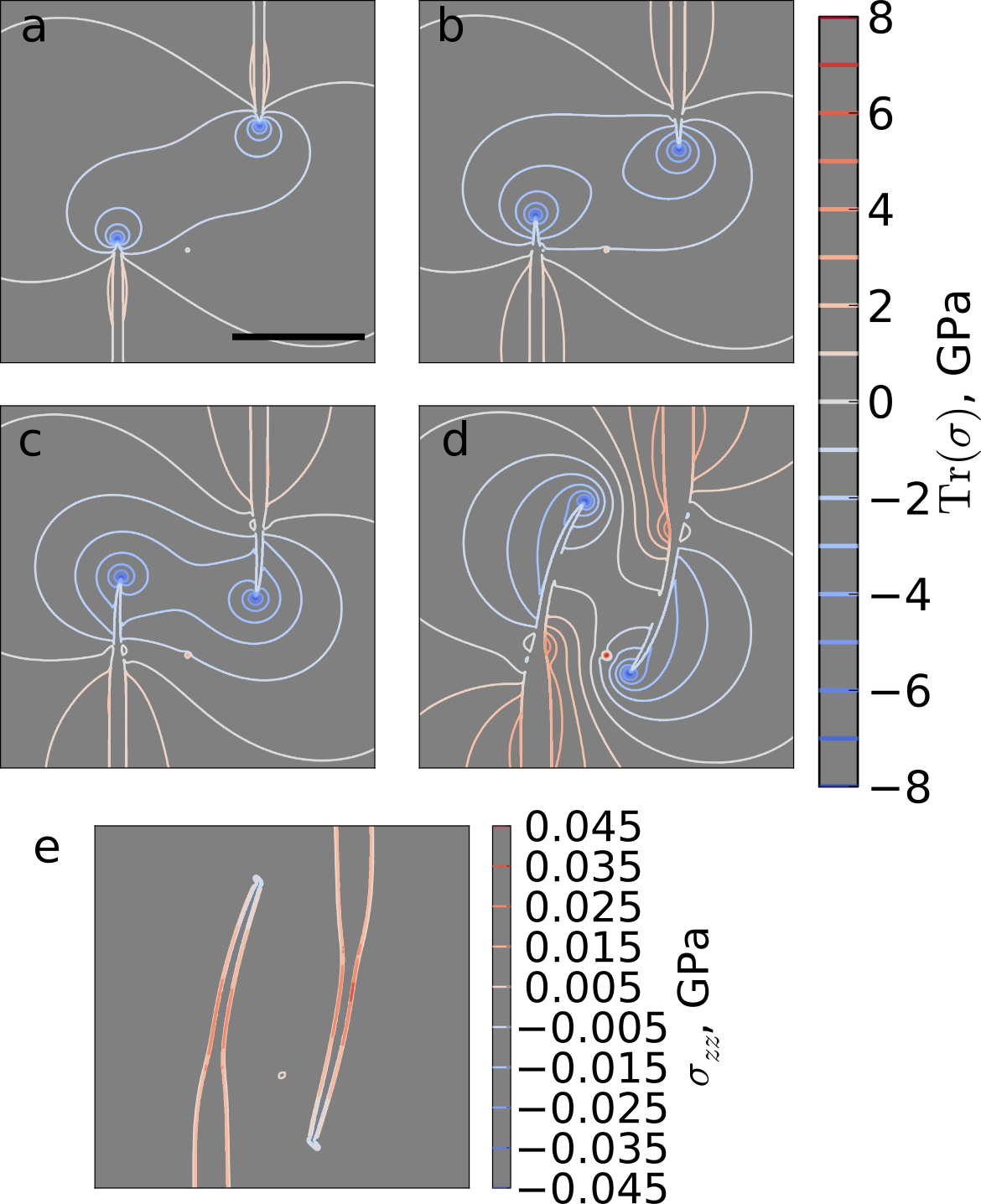} 
\caption{\label{wrinkle_pair_stress} (a--d) Contour lines of the stress trace during the evolution of an avoiding pair of wrinkles, with a field of view that excludes the sheet edges. The nucleus separations are $X=Y=56.6$ nm.  The line joining the wrinkle tips is a local maximum in the stress distribution and the tips propagate around the centre of this line, which is a saddle point. 
(e) Contour lines of $\sigma_{zz}$, for the same wrinkle configuration as shown in panel (d).
For clarity,  the interiors of the wrinkles (defined by a height threshold of 2 \AA) are not shown. All images have the same scale; the scale bar, shown in panel (a), represents $100$ nm.}
\end{figure}

When two wrinkles are grown in proximity to each other, their behavior is determined by the superposition of their stress fields. A video showing the evolution of stresses for a growing pair of wrinkles is provided as Supplemental Materials~\cite{SuppMat}. As shown in Fig.~\ref{wrinkle_pair_stress}, wrinkles propagate along the steepest stress gradient in front of them.  Pairs of wrinkles propagate in a spiral around the centre of the line joining their tips. 
As seen in Fig.~\ref{wrinkle_pair_stress}(a,b),  the wrinkle tips initially propagate outwards slightly, and the high-stress central region expands. 
As the tips pass each other, their motion changes to an inwards propagation, as seen in Fig.~\ref{wrinkle_pair_stress}(c), and the high-stress region contracts. The wrinkles continue to propagate around the centre of the line joining their tips, 
and come to a stop in the configuration seen in Fig.~\ref{wrinkle_pair_stress}(d). This configuration is stabilized by the balance between applied forces and stresses in the sheet. However, even for larger upward forces on the wrinkle nuclei, the deformations required to merge two wrinkles are highly unfavourable. This is consistent with the observation of stable avoiding pairs of wrinkles in simulation~\cite{Zhu2014} and experimental~\cite{Yamamoto2012} studies of graphene deposited on nanoparticles. The dependence of wrinkle extension on nucleus tip distance seen in Fig.~\ref{wrinkle_shape}(c) can be accounted for by noting that when nuclei are close, wrinkle propagation is limited by wrinkles blocking each other, but as nuclei distance increases, stresses decrease and the rate of wrinkle propagation decreases accordingly.

Finally, we observe that, as expected for a thin plate~\cite{LandauLifshitz}, the stresses induced by wrinkles act in-plane. As shown in Fig.~\ref{wrinkle_pair_stress}(e), $\sigma_{zz}$ is zero outside the wrinkle, and even on the wrinkle boundary has a value two orders of magnitude smaller than $\sigma_{xx}+\sigma_{yy}$. In other words, wrinkle growth is driven by buckling induced by in-plane compressive stresses.

\subsection{Delamination transition for $n$-layer graphene suspended on four particles}
\begin{figure}
\includegraphics[width=\columnwidth]{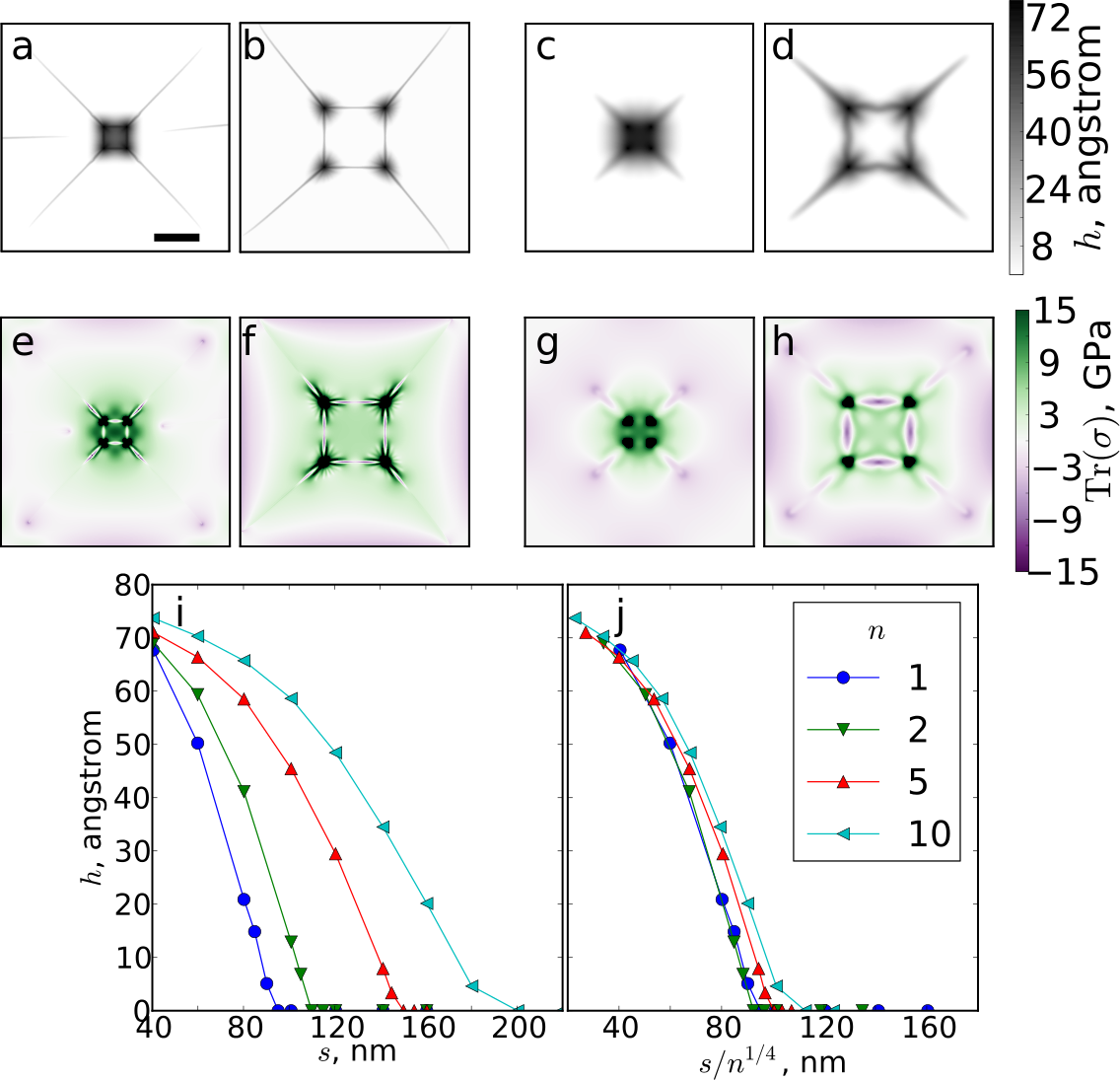} 
\caption{\label{four_particles} Typical configurations of a sheet deposited on four nanoparticles: (a) monolayer sheet, particle spacing $60$ nm, (b) monolayer sheet, particle spacing $160$ nm, (c) 5-layer sheet, particle spacing $60$ nm, (d) 5-layer sheet, particle spacing $160$ nm. Panels (e--h) show the trace of the stress tensor corresponding to the configurations. Where the stress is above $15$ GPa, the color scale has been clipped (black regions on top of the particles).  All eight panels have the same scale, and the scale bar, shown in panel (a), represents $100$ nm. (i) Height of the center of the graphene sheet as a function of particle spacing, for 1, 2, 5 and 10 layer sheets. (j) As a first approximation, the delamination transition occurs at a particle spacing $s^*\sim n^{1/4}$.}
\end{figure}

Finally, we consider a situation where wrinkle merging is of great relevance, namely the transition from conformation to delamination of a graphene sheet placed on a substrate decorated with nanoparticles.
To study this transition in a systematic way, we simulate a $600\times600$~nm$^2$ graphene sheet deposited on a substrate with four nanoparticles of diameter $8$ nm, arranged in a rectangle with aspect ratio close to unity and a range of side lengths $\ell$, with center close to the sheet center. (We use this configuration, rather than a square centered exactly, in order to avoid spurious effects arising from symmetry.) Example movies of deposition are shown in the Supplemental Material~\cite{SuppMat}. Here, we examine the role of both the inter-particle distance and the sheet's mechanical properties, which are controlled by its thickness.

Figure~\ref{four_particles}(a--d) shows typical configurations for close-spaced and distant particles, for a monolayer (panels a and b) and a 5-layer sheet (panels c and d). When the particles are close together, the sheet is delaminated in the center of the rectangle, but as their spacing is increased, it is able to conform to the substrate. Increasing the sheet thickness ``smooths'' the sheet and increases the size of features, as can be seen by comparing panels (a,b) and (c,d) of Fig.~\ref{four_particles}. It also reduces stresses in the sheet, as seen in panels (e--h) which show the trace of the stress tensor. Near the particles, stresses are tensile ($\mathrm{Tr}(\sigma)>0$), but compressive regions exist inside the wrinkles and especially at the wrinkle tips. The effect of the finite sheet size is also seen, with the stress fields clearly modified by the sheet edges.

The delamination transition is quantified in Fig.~\ref{four_particles}(i), which shows the height of the center of the sheet as a function of particle spacing $s$, for 1-, 2-, 5- and 10-layer sheets. In all cases, a sharp transition is observed at a particle spacing $s^*$ that depends on sheet thickness. 
As a first approximation, one can estimate the dependence on sheet thickness under the assumption that bending rigidity does not enter the problem and the sheet deforms purely via stretching. In that case the delaminated region of an $n$-layer sheet around a protrusion has radius $R \approx r (4 n E_{2D}/3\Gamma)^{1/4}$~\cite{Zong2010,Yamamoto2012}, where $r$ is the particle radius, $E_{2D}$ is the tensile rigidity of the sheet and $\Gamma$ characterizes the adhesion energy. Under the assumption that delamination in the center occurs when delaminated regions meet, the critical particle spacing scales as $s^*\sim n^{1/4}$. As seen in Fig.~\ref{four_particles}(j), this scaling collapses the 1- and 2-layer curves very well, but already breaks down for the 5-layer sheet. This is to be expected, because argument does not take into account the existence of long-range stresses and wrinkles connecting the particles.


\section{Discussion}

\subsection{Comparison of interacting wrinkles with interacting cracks}
Delamination is a form of fracture, so it is not surprising that our interacting delamination wrinkles are reminiscent of interacting cracks. 
Single wrinkles and cracks have different fracture modes, since propagation of a crack due to pulling orthogonal to the crack direction is purely mode I fracture, whereas wrinkle propagation is inherently a mixture of modes I and II, due to the compressive stresses in front of the wrinkle tip (as shown in Fig.~\ref{single_wrinkle}). Nonetheless, when interactions with other cracks or wrinkles are introduced, both types of fracture propagation involve mixed modes I and II loading, and in both cases fracture proceeds as rotation of the line connecting their tips, which is a local stress maximum~\cite{Fender2010}, tensile for cracks and compressive for wrinkles.

Experiments on interacting cracks in gelatine plates~\cite{Fender2010} reveal a universal shape for the lenticular region cut out by two interacting cracks, described by $y(x)=\sqrt{x}$. In those experiments, notch lateral separation (equivalent to our $X$, see Fig.~\ref{wrinkle_shape}(a)) was varied over an order of magnitude, with fixed separation along the direction of the notches (equivalent to our $Y$ in Fig.~\ref{wrinkle_shape}(a)). By way of contrast, we are unable to identify a universal shape in our data on avoiding wrinkle pairs, as is obvious from visual inspection of Fig.~\ref{wrinkle_shape}(c), in which some wrinkles are almost straight and others are substantially curved. Furthermore, in those experiments, cracks were found to only deviate from straight paths when their tips pass each other, whereas we find wrinkle shapes that are curved even at the wrinkle nuclei. We attribute these differences in behavior to differences in the long range stress fields. In addition to fundamental differences between crack and wrinkle tips, in our simulations stresses are strongly affected by system edges, as we discuss below.

Experiments by Cortet {\it et al} on cracks in paper also reveal crack shapes controlled by interactions \cite{Cortet2008}. Their experimental geometry is more complex, consisting of two lines of equidistant slits. The cracks are grown by pulling the paper perpendicular to the slit direction. The slits and their spacing are both fixed, and the control parameter is the distance $d$ between the two lines. 
For small $d$, cracks typically repel each other during their initial growth, before becoming attracted. For larger $d$, repulsion is seen less frequently, until a transition inter-line distance above which cracks from different lines do not interact. As seen in the wrinkle shape data for $X/Y=1$ in Fig.~\ref{wrinkle_shape}(c), this behavior is borne out by our wrinkles, however we have deterministic rather than probabilistic wrinkle propagation due to a lack of disorder in our systems and a simplified geometry with only two wrinkles rather than an array of cracks.

Finally, we note a fundamental geometrical difference between the wrinkle and crack scenarios. Wrinkles can move laterally without broadening, whereas in the absence of healing cracks can only grow at their tips or by opening up. In this sense, crack propagation is ``irreversible'', whereas wrinkles can move around to reduce stresses, as seen in example wrinkle propagation movies in Supplemental Material~\cite{SuppMat}. This introduces a subtlety into the analysis of wrinkle shapes, because their final shape is not necessarily a trace of the motion of their tips.

\subsection{Open boundary conditions and edge effects}
We now discuss the significance of open boundary conditions in our simulations. As observed in Fig.~\ref{single_wrinkle}, in an infinite sheet the stresses generated by a single wrinkle decay as a power law with distance, whereas the measured $\mathrm{Tr}(\sigma)$ from a single wrinkle in our coarse grained simulations has a much faster decay, being cut off at the sheet edges. The stresses in a sheet deposited on four nanoparticles, shown in Fig.~\ref{four_particles}, also exhibit clear edge effects. 
While the presence of edge effects in our simulations does have drawbacks in terms of directly applying our quantitative results to larger systems, they are in fact necessary for qualitative agreement of wrinkle patterns with experiments.

\begin{figure}[htp]
\centering
\includegraphics[width=\columnwidth]{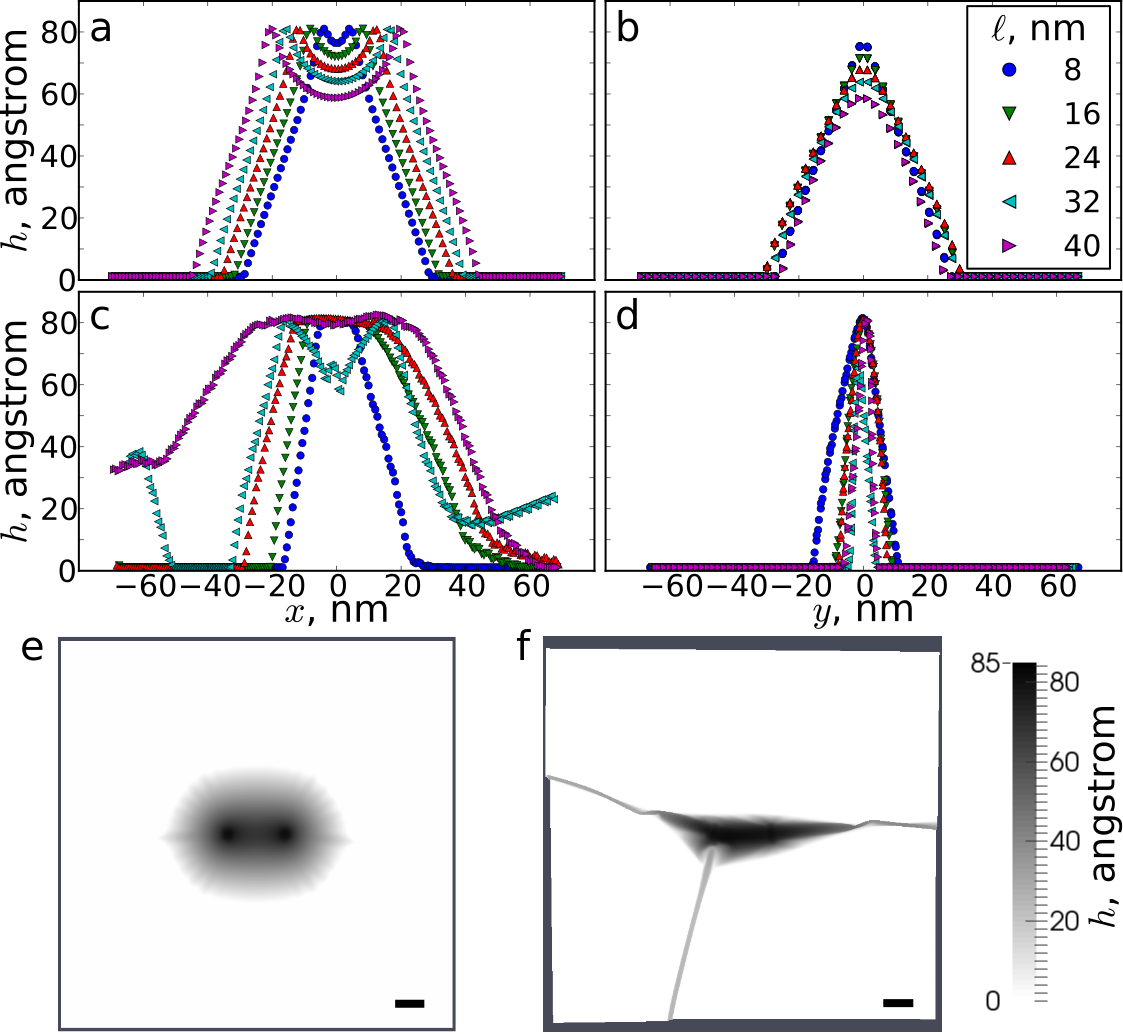} 
\caption{\label{atomistic_lj_bcs} Wrinkle profiles in atomistic simulations, (a) along the $x$ direction and (b) along the $y$ direction, with periodic boundary conditions. (c,d) Corresponding profiles for simulations with open boundary conditions. The graphene sheet is modeled using an AIREBO potential and deposited on a substrate with which it interacts {\it via} a radial Lennard-Jones potential. The substrate has spherical protrusions of diameter $8$ nm which are placed on the $x$ axis (which is parallel to the armchair direction of the graphene) with separation $\ell$ as indicated in the legend. The lower panels show height maps for configurations obtained with $\ell=20$ nm, for (e) periodic and (f) open boundary conditions. In panel (f), the contraction of the sheet at the edges can be seen. Scale bars represent $10$ nm.}
\end{figure}

To see this, we compare simulations of an atomistic $140\times140$ nm$^2$ sheet of monolayer graphene interacting with a substrate with two nanoparticles of diameter $8$ nm via a radial Lennard-Jones potential, with open and periodic boundary conditions. Details of the simulations are given in Appendix~\ref{atomistic}, and sample movies are provided as Supplemental Material~\cite{SuppMat}. As seen in Fig.~\ref{atomistic_lj_bcs}, periodic boundary conditions result in a very smooth deformation of the sheet, which remains flat and attached to the substrate away from the particle pair. We attribute this to the fact that the total dimensions of the sheet are fixed. This places a strong constraint on out of plane buckling which must, in this case, involve costly stretching. On the other hand, open boundary conditions allow wrinkles to form not only as a delaminated region between the particles, but also away from the particles, in a manner reminiscent of that seen in the experiments of Yamamoto {\it et al}~\cite{Yamamoto2012}.

The importance of open boundary conditions also raises questions about a recent simulation study of delamination wrinkles connecting pairs of nanoparticles~\cite{Zhu2014}, since that study used periodic boundary conditions along the direction of the nanoparticle center-center line. In particular, the authors find a transition between avoiding and joined wrinkles depending on particle size and spacing, but our results suggest this transition would be quantitatively different if open boundary conditions were used.

\subsection{Importance of lateral deformation}
\begin{figure}
\includegraphics[width=\columnwidth]{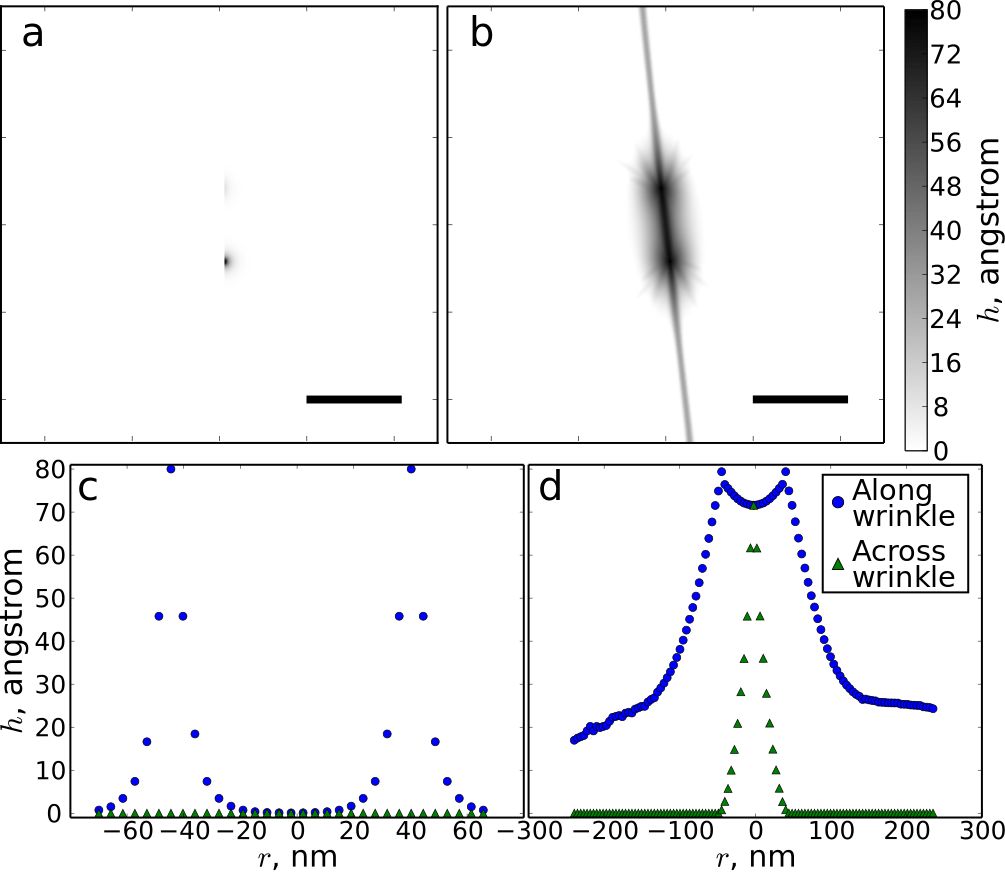} 
\caption{\label{scalar_model_bad} Height maps of a monolayer graphene sheet simulated using (a) the scalar model and (b) the coarse grained model, on a substrate with two spherical protrusions of diameter $8$ nm and centre-centre separation $84.6$ nm. (c,d) Corresponding height profiles along lines parallel and perpendicular to the protrusion centre-centre line. In the scalar model (panel c), the sheet is delaminated only in a circular region close to the protrusions, whereas the triangulated sheet (panel d) supports extended wrinkles. Scale bars indicate $100$ nm.}
\end{figure}

The necessity of open boundary conditions for realistic results also points to the importance of allowing in-plane modes in the graphene sheet in the scenarios we study here. To demonstrate this, we have studied a model based on random fuse models commonly used to study fracture~\cite{deArcangelis1985, Duxbury1987}, consisting of a lattice of sites $i$ with height $h_i$, with forces on each site arising from adhesion, bending and stretching as described in Appendix~\ref{fuse_model}. The key point of the model is that each site has fixed position in plane, that is, the sheet deforms only through vertical motion. 

As seen in Fig.~\ref{scalar_model_bad}, the fuse model does not yield a realistic description of graphene on a patterned substrate and cannot describe extended wrinkles. Although some delamination does occur around the particles, as expected~\cite{ViolaKusminskiy2011}, the delaminated regions do not interact except to form a linear superposition if they overlap. This is in stark contrast to both experiments and our more realistic model.  
Models in which the sheet is fully characterized by a scalar height field are common in the literature on graphene~\cite{PierreLouis2008, Cullen2010, Li2010, NeekAmal2010, Wagner2012, Yamamoto2012}.
Although such models offer computational efficiency and even the possibility of analytical progress, at least in situations with a high level of symmetry~\cite{ViolaKusminskiy2011}, in the present scenario of graphene wrinkling on patterned substrates these models are not appropriate because they overlook essential physics.

\begin{figure}
\includegraphics[width=\columnwidth]{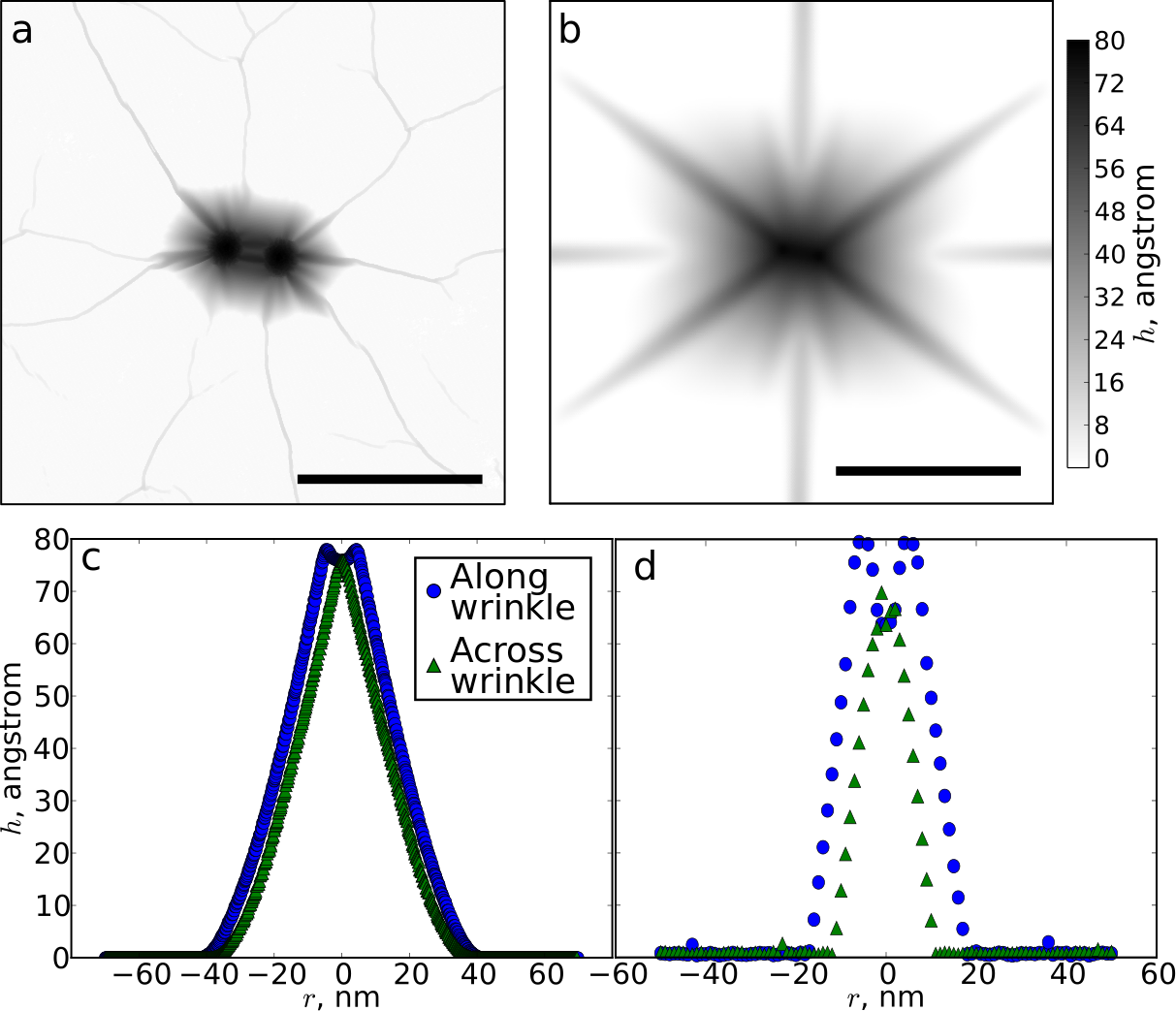} 
\caption{\label{too_strong_substrate} Height maps of a $140\times140$ nm$^2$ monolayer graphene sheet simulated  (a) atomistically (using an AIREBO potential), with an atomistic Si substrate with a Tersoff interaction between sheet and substrate and (b) using the coarse grained model. In both cases, substrate has two spherical protrusions of diameter $8$ nm and center-center separation $10$ nm, with the center-center line at $10^{\circ}$ to the $x$ axis (the graphene armchair direction in the atomistic simulations) to ensure the configurations are not affected by spurious lattice effects. (c,d) Corresponding height profiles along lines parallel and perpendicular to the protrusion centre-centre line. Scale bars indicate $50$ nm.}
\end{figure}

The need for realistic in-plane deformations is also relevant for fully-atomistic simulations, in which both the sheet and the substrate are simulated atomstically. An example configuration obtained in such simulations of graphene deposited on a substrate with two nanoparticles is shown in Fig.~\ref{too_strong_substrate}, with a configuration from the same scenario simulated using our coarse grained model also shown for comparison. In these atomistic simulations, the substrate is crystalline silicon, and interacts with the graphene {\it via} a Tersoff interaction~\cite{Tersoff1988}, which is rather strong. We find the graphene adheres strongly on contact and once adhered, can neither detach nor move laterally. As a result, although wrinkles are present in the graphene, they are lower than those observed experimentally and are atomically narrow.  We believe they correspond to boundaries between regions of graphene that have adhered to the substrate in an incompatible way, and are therefore fundamentally different in origin to the wrinkles we observe in our model.

\section{Conclusions}
We have presented a systematic study of the interactions and coalescence of delamination patterns in thin sheets, as exemplified by wrinkles in graphene on patterned substrates. We have shown how the morphology of pairs of wrinkles is determined by their interactions, which are mediated by long range stress fields. We have also studied a delamination transition driven by wrinkle coalescence in graphene suspended over a group of nanoparticles, finding rich and complex behavior controlled by the particle positions, sheet thickness and finite size. These results reveal the importance of sheet edges, which not only modify stress fields but enable in-plane deformation of the sheet, without which it is not possible to replicate experimental results.

Our results are important for controlling the morphology of graphene sheets for applications, but are not specific to graphene and can be applied also to other thin films. As demonstrated in Table~\ref{wrinkles_in_literature} and discussed in the Introduction, avoiding pairs of wrinkles are a universal feature that can be found even on the lengthscales of geomembranes. Accordingly, we expect our results to be experimentally testable not only in graphene but in other thin sheets, so that tabletop experiments would be possible. 

Finally, we note that our simulations of graphene on groups of particles point to the possibility of a unified framework for understanding delamination due to substrate pillars/particles, in which groups of pillars form ``delamination building blocks'' where delamination occurs locally. One such block is that highlighted in Fig.~\ref{yamamoto_comparison}(f,g), where a monolayer sheet conforms but a 5-layer sheet delaminates. In this picture, global delamination arises arises from the complex interplay between the random arrangement and the elastic interactions of these building blocks.

\section*{Acknowledgements}
Helpful discussions with Michael Zaiser are gratefully acknowledged.
This work is supported by the European Research Council through the Advanced Grant 2011 SIZEFFECTS. 

\appendix

\section{Implementation details for coarse grained model}
\label{lammps}
We implement our simulations using the molecular dynamics package LAMMPS~\cite{Plimpton1995}. Particles are initially generated in a flat triangular lattice. They are connected by harmonic bonds using the {\tt bond\_style harmonic} command, to give a stretching cost to the sheet, as described in the main text. The bending cost of the sheet is implemented using improper interactions between sets of four particles on the vertices of two neighboring triangles, using the {\tt improper\_style harmonic} command. 

The sheet is pushed downward with a constant position-independent force of $0.001$ eV/\AA. The substrate consists of a flat repulsive sheet, which is implemented as a {\tt plane} which interacts with the particles through a {\tt wall/region} fix, with a harmonic interaction used. The interaction has spring constant $1.0$ eV/\AA$^2$ and a cutoff distance of $1.0$ \AA, with the energy minimum on the substrate surface. Only the repulsive part of the harmonic potential is used, and we have found similar results using a Lennard-Jones interaction. The nanoparticles are implemented using the {\tt indent} fix (spring constant $1.0$ eV/\AA$^2$), which allows the nanoparticles to deform slightly under the influence of the sheet. This avoids errors in which the particles of the sheet intersect with the nanoparticle region, which we found to be a problem when using a {\tt wall/region} fix for the nanoparticles. 

We implement friction with the substrate by defining a {\tt group} of edge atoms at the simulation start, and using a {\tt spring/self} fix to couple the atoms to their initial lateral positions using springs of stiffness  of stiffness $0.001$ eV/\AA$^2$ (no vertical forces are applied by this fix). The edge region typically consists of the atoms outside a square with side length $\sim 10$ nm less than the sheet side length, but we do not see any sensitivity to the details of this except in very small sheets. We drain energy from the system with a {\tt viscous} fix with damping parameter $5\times 10^{-4}$ eV/ps.  We run our simulations until no change in sheet configuration can be seen in two snapshots separated by $10^4$ time steps. We use a time step of $0.0005$ ps.

\label{stress_calc}
Stresses are computed using the built-in {\tt stress/atom} command. For our system, this returns for particle $i$ a tensor
\begin{equation}
\begin{split}
\label{stress}
S_{\alpha \beta}^{(i)} = - \biggl [ m v_{\alpha}^{(i)} v_{\beta}^{(i)} 
	+ \frac{1}{2} (N_B^{(i)} r_{\alpha}^{(i)} F_{\beta}^{(i)} + \sum_{n=1}^{N_B^{(i)}} r_{\alpha}^{(n)} F_{\beta}^{(n)})\\ 
+ \frac{1}{4}( N_I^{(i)} r_{\alpha}^{(i)} F_{\beta}^{(i)} + \sum_{n=1}^{N_I^{(i)}} \sum_{j=1}^3 r_{\alpha}^{(j)} F_{\beta}^{(j)}) \biggr].
\end{split}
\end{equation}
The first term is a contribution from the kinetic energy, which is zero at the end of our simulations, when we measure the stress. The second term gives contributions from each of the $N_B^{(i)}$ bonds particle $i$ is a member of, and the final term gives contributions from the $N_I^{(i)}$ impropers particle $i$ is a member of. $r^{(i)}$ is the position of particle $i$ and $F^{(i)}$ refers to the force on $i$ as a result of the relevant interaction. The quantity $S$ given by Eq.~\eqref{stress} is not strictly a stress, but is multiplied by a characteristic volume and is returned with units bars \AA$^2$. We obtain the stress $\sigma$ on the sheet by taking the particle volume to be the area of a triangle (here $12.5$ \AA$^2$), multiplied by the sheet thickness, which is approximately $3n$ \AA. We also note that we focus on the stress trace in part because it is invariant so that we don't need to consider the local orientation of the sheet.

\section{Simplified scalar model}
\label{fuse_model}
Our model can be greatly simplified by fixing the particle positions laterally, so that the sheet can be described entirely by a height function $u(x,y)$. In this case, $u$ is given by the solution to
\begin{equation}
\label{fuse_model_eq}
B\nabla^4 u - E \nabla^2 u + f(u-s)=0,
\end{equation}
where $B$ is the bending rigidity, $E$ is Young's modulus, and $f$ is a function describing the interaction of the sheet with a substrate of height $s(x,y)$. We use a piecewise $f$,
\begin{equation}
\label{fuse_attachments}
f(z) = \left\{
        \begin{array}{ll}
            -\kappa z & \quad z \leq 0, \\
            k z & \quad z^* \geq z > 0,\\
            0 & \quad z > z^*,
        \end{array}
    \right.
\end{equation}
where $\kappa \gg k$ to model a hard repulsion from the substrate and $z^*$ is the delamination threshold.

To solve \eqref{fuse_model_eq}, we discretize it onto a square lattice with lattice constant $5$ \AA. In analogy to fuse models \cite{deArcangelis1985, Duxbury1987} for fracture, we solve iteratively, which can be done efficiently using the algorithm proposed in Ref.~\cite{Nukala2003}. The discretized form
of Eq.~\eqref{fuse_model_eq} can be written in matrix form
\begin{equation}
\label{fuse_matrix}
A_{ij}z_i=f_j,
\end{equation}
where $\mathbf{A}$ is the discrete form of $B\nabla^4-E\nabla^2$, $z_i$ is the height of the sheet at site $i$ and $\mathbf{f}$ is the the vector of forces on the sheet induced by the nanoparticles or substrate. After solving the matrix equation~\eqref{fuse_matrix}, we update the forces to allow for delamination or adhesion of the sheet, in correspondence with the forces prescribed in Eq~\ref{fuse_attachments}. We then recalculate the height distribution $\mathbf{z}$, and repeat the process until the system reaches an equilibrium where nothing changes. This method implicitly assumes that the stress relaxation in the lattice system is much faster than the delamination of a site and therefore the stress in the system is redistributed instantaneously.

In the results presented in Fig.~\ref{scalar_model_bad}, we use parameter values $B=1$~eV, $E=2\times10^3$~eV/nm$^2$, $k=5$~eV/nm$^2$ (Ref.~\cite{Koenig2011}), $\kappa=10^4$~eV/nm$^4$, $z^*=0$. The choice $z^*=0$ corresponds to the limit of adhesion forces only acting on contact and should give maximal delamination. We have also tested finite $z^*$, as well as stronger and weaker adhesion forces, and find similar results.

\section{Atomistic simulations}
\label{atomistic}
We performed molecular dynamics simulations of a single layer, monocrystalline graphene patch composed of approximately 570,000 atoms, interacting with a substrate. The patch has approximate dimensions \(140 \times 140\)~nm$^2$.
All simulations were performed using the LAMMPS molecular dynamics toolbox~\cite{Plimpton1995}.
Carbon-carbon atom interaction was modeled using an atomistic pairwise potential. We used the ``Adaptive Intermolecular REactive Bond Order'' (AIREBO) potential~\cite{Stuart2000}, which was originally developed as an extension of the ``REactive Bond Order'' potential (REBO)~\cite{brenner2002second}.
The AIREBO potential includes an adaptive treatment of non-bonded and dihedral angle interactions that is employed to capture the bond breaking and bond reformation between carbon atoms, offering a valid tradeoff between simulation accuracy and computational efficiency; it has been extensively used to simulate and predict mechanical properties of carbon-based materials, i.e. fullerenes, carbon nanotubes and graphene~\cite{Zhao2009}.
The graphene layer is thermalized using a Berendsen~\cite{Berendsen1984} thermostat set at 300K, with a characteristic relaxation time set to 0.1~ps; the simulation timestep is set at 1 fs to ensure a correct time integration of the atom dynamics.

The geometry of the substrate consists of a flat plane with two spherical particles on top.
The interaction between graphene and substrate was simulated in two different ways: model A, using a continuum-like perpendicular wall interaction; model B, using an atomistic substrate.
Model A defines the substrate as a (continuous) wall that interacts with the carbon atoms by generating a force in a direction perpendicular to the wall. The modulus of the force of the wall-particle interactions is computed from the Lennard-Jones potential:
\( V(r)=4 \epsilon \left[ \left( \frac{\sigma}{r} \right)^{12} - \left(\frac{\sigma}{r} \right)^6 \right] \).
The parameters \(\epsilon=0.04\)~eV, \(\sigma=0.1\)~nm, and the wall-atom long-range cutoff distance (\(d=2.5\)~nm) were chosen to match the {C}--{SiO}$_{2}$ interaction, as indicated in~\cite{Yamamoto2012}.
Since this interaction generates forces perpendicular to the substrate, model A represents an idealized case of frictionless contact.
Model B has the same geometrical features of model A, but in this case the substrate is composed of a Si monocrystal consisting of approximately 3~million atoms. The pairwise atomic interactions are hybrid: C-C interaction uses the AIREBO potential, while C-Si interaction is defined by the ``Tersoff" pairwise potential, as described in Ref.~\cite{Tersoff1988}. In this case, the atoms of the substrate are not time-integrated (they remain frozen).
Due to the strong coupling between C and Si atoms, model B represents a limit case in which the graphene layer encounters a very large resistance to lateral slip.

The simulation protocol consists of four steps.
(i) the graphene layer is generated, and the system is brought to the energy minimum (we used the standard LAMMPS ``conjugate gradient algorithm" (CG) minimization);
(ii) the graphene layer is placed above the substrate, just beyond the potential cutoff distance, and the atoms are given a small initial speed directed towards the substrate.
(iii) the graphene interacts with the substrate and becomes attached to it; the simulation is run until the whole patch has come in contact; in some cases (PBC), periodic oscillations around an equilibrium configuration can be observed; the thermostat induces only minimal damping, and the oscillations can be observed for long times.
(iv) we drive the system to its equilibrium configuration by removing the thermostat and by adding an artificial viscous damping term. The simulation is stopped once a lower threshold in atom motion is reached.

\section{Stresses induced by an idealized wrinkle in an infinite sheet}
\label{FvK}
For analytical tractability, we work under the assumption that the sheet deforms only out of plane. In our idealized calculation, we impose a wrinkle defined by the height function $u$
\begin{equation}
\label{analytical_wrinkle_shape}
u(x,y) = \exp(-(x/w)^2)\bigl(1-\mathrm{erf}(\sqrt{\pi}y)\bigr),
\end{equation}
that is, with a Gaussian profile along its cross section with width $w$ (which we set to $2$ nm) and extending to $-\infty$ along the $y$ axis. Following standard methods~\cite{LandauLifshitz}, the stresses in an infinite sheet are obtained from the auxiliary function $\chi$:
\begin{equation}
\sigma_{xx} = \frac{\partial^2 \chi}{\partial y^2},\qquad
\sigma_{yy} = \frac{\partial^2 \chi}{\partial x^2},\qquad
\sigma_{xy} = \frac{\partial^2 \chi}{\partial x \partial y},
\end{equation}
which satisfies
\begin{equation}
\nabla^2 \chi + E \bigl( \frac{\partial^2 u}{\partial x^2} \frac{\partial^2 u}{\partial y^2} - (\frac{\partial^2 u}{\partial x \partial y})^2 \bigr) = 0,
\end{equation}
where $E=2.4$ TPa (Ref.~\cite{Lee2012, Hartmann2013}) is Young's modulus. The numerical solution for $\sigma_{xx}+\sigma_{yy}$ (which is the trace of the in-plane stress for a sheet parallel to the $xy$ plane) is shown in Fig.~\ref{single_wrinkle}(c,d).


\begin{thebibliography}{10}%
\makeatletter
\providecommand \@ifxundefined [1]{%
 \ifx #1\undefined \expandafter \@firstoftwo
 \else \expandafter \@secondoftwo
\fi
}%
\providecommand \@ifnum [1]{%
 \ifnum #1\expandafter \@firstoftwo
 \else \expandafter \@secondoftwo
\fi
}%
\providecommand \enquote [1]{``#1''}%
\providecommand \bibnamefont  [1]{#1}%
\providecommand \bibfnamefont [1]{#1}%
\providecommand \citenamefont [1]{#1}%
\providecommand\href[0]{\@sanitize\@href}%
\providecommand\@href[1]{\endgroup\@@startlink{#1}\endgroup\@@href}%
\providecommand\@@href[1]{#1\@@endlink}%
\providecommand \@sanitize [0]{\begingroup\catcode`\&12\catcode`\#12\relax}%
\@ifxundefined \pdfoutput {\@firstoftwo}{%
 \@ifnum{\z@=\pdfoutput}{\@firstoftwo}{\@secondoftwo}%
}{%
 \providecommand\@@startlink[1]{\leavevmode}%
 \providecommand\@@endlink[0]{}%
}{%
 \providecommand\@@startlink[1]{%
  \leavevmode
  \pdfstartlink
   attr{/Border[0 0 1 ]/H/I/C[0 1 1]}%
   user{/Subtype/Link/A<</Type/Action/S/URI/URI(#1)>>}%
  \relax
 }%
 \providecommand\@@endlink[0]{\pdfendlink}%
}%
\providecommand \url  [0]{\begingroup\@sanitize \@url }%
\providecommand \@url [1]{\endgroup\@href {#1}{\urlprefix}}%
\providecommand \urlprefix [0]{URL }%
\providecommand \Eprint[0]{\href }%
\@ifxundefined \urlstyle {%
  \providecommand \doi [1]{doi:\discretionary{}{}{}#1}%
}{%
  \providecommand \doi [0]{doi:\discretionary{}{}{}\begingroup
  \urlstyle{rm}\Url }%
}%
\providecommand \doibase [0]{http://dx.doi.org/}%
\providecommand \Doi[1]{\href{\doibase#1}}%
\providecommand \bibAnnote [3]{%
  \BibitemShut{#1}%
  \begin{quotation}\noindent
    \textsc{Key:}\ #2\\\textsc{Annotation:}\ #3%
  \end{quotation}%
}%
\providecommand \bibAnnoteFile [2]{%
  \IfFileExists{#2}{\bibAnnote {#1} {#2} {\input{#2}}}{}%
}%
\providecommand \typeout [0]{\immediate \write \m@ne }%
\providecommand \selectlanguage [0]{\@gobble}%
\providecommand \bibinfo [0]{\@secondoftwo}%
\providecommand \bibfield [0]{\@secondoftwo}%
\providecommand \translation [1]{[#1]}%
\providecommand \BibitemOpen[0]{}%
\providecommand \bibitemStop [0]{}%
\providecommand \bibitemNoStop [0]{.\EOS\space}%
\providecommand \EOS [0]{\spacefactor3000\relax}%
\providecommand \BibitemShut [1]{\csname bibitem#1\endcsname}%
\bibitem{Sun2006}%
  \BibitemOpen
  \bibfield{author}{%
  \bibinfo {author} {\bibfnamefont{Yugang}\ \bibnamefont{Sun}}, \bibinfo
  {author} {\bibfnamefont{Won~Mook}\ \bibnamefont{Choi}}, \bibinfo {author}
  {\bibfnamefont{Hanqing}\ \bibnamefont{Jiang}}, \bibinfo {author}
  {\bibfnamefont{Yonggang~Y.}\ \bibnamefont{Huang}},\ and\ \bibinfo {author}
  {\bibfnamefont{John~A.}\ \bibnamefont{Rogers}},\ }%
  \bibfield{title}{%
  \enquote{\bibinfo {title} {Controlled buckling of semiconductor nanoribbons
  for stretchable electronics},}\ }%
  \bibfield{journal}{%
  {\bibinfo {journal} {Nature Nanotechnology}}\ }%
  \textbf{\bibinfo {volume} {1}},\ \bibinfo {pages} {201} (\bibinfo {year}
  {2006})%
  \bibAnnoteFile{NoStop}{Sun2006}%
\bibitem{Malachias2008}%
  \BibitemOpen
  \bibfield{author}{%
  \bibinfo {author} {\bibfnamefont{Angelo}\ \bibnamefont{Malachias}}, \bibinfo
  {author} {\bibfnamefont{Yongfeng}\ \bibnamefont{Mei}}, \bibinfo {author}
  {\bibfnamefont{Ratna~K.}\ \bibnamefont{Annabattula}}, \bibinfo {author}
  {\bibfnamefont{Christoph}\ \bibnamefont{Deneke}}, \bibinfo {author}
  {\bibfnamefont{Patrick~R.}\ \bibnamefont{Onck}},\ and\ \bibinfo {author}
  {\bibfnamefont{Oliver~G.}\ \bibnamefont{Schmidt}},\ }%
  \bibfield{title}{%
  \enquote{\bibinfo {title} {Wrinkled-up nanochannel networks: Long-range
  ordering, scalability, and x-ray investigation},}\ }%
  \bibfield{journal}{%
  \Doi{10.1021/nn800308p}{\bibinfo {journal} {ACS Nano}}\ }%
  \textbf{\bibinfo {volume} {2}},\ \bibinfo {pages} {1715--1721} (\bibinfo
  {year} {2008})%
  \bibAnnoteFile{NoStop}{Malachias2008}%
\bibitem{Mei2010}%
  \BibitemOpen
  \bibfield{author}{%
  \bibinfo {author} {\bibfnamefont{Yongfeng}\ \bibnamefont{Mei}}, \bibinfo
  {author} {\bibfnamefont{Suwit}\ \bibnamefont{Kiravittaya}}, \bibinfo {author}
  {\bibfnamefont{Stefan}\ \bibnamefont{Harazim}},\ and\ \bibinfo {author}
  {\bibfnamefont{Oliver~G.}\ \bibnamefont{Schmidt}},\ }%
  \bibfield{title}{%
  \enquote{\bibinfo {title} {Principles and applications of micro and nanoscale
  wrinkles},}\ }%
  \bibfield{journal}{%
 {\bibinfo {journal}
  {Materials Science and Engineering: R: Reports}}\ }%
  \textbf{\bibinfo {volume} {70}},\ \bibinfo {pages} {209 -- 224} (\bibinfo
  {year} {2010})%
  \bibAnnoteFile{NoStop}{Mei2010}%
\bibitem{Audoly1999}%
  \BibitemOpen
  \bibfield{author}{%
  \bibinfo {author} {\bibfnamefont{Basile}\ \bibnamefont{Audoly}},\ }%
  \bibfield{title}{%
  \enquote{\bibinfo {title} {Stability of straight delamination blisters},}\ }%
  \bibfield{journal}{%
  \Doi{10.1103/PhysRevLett.83.4124}{\bibinfo {journal} {Phys. Rev. Lett.}}\ }%
  \textbf{\bibinfo {volume} {83}},\ \bibinfo {pages} {4124--4127} (\bibinfo {year} {1999})%
  \bibAnnoteFile{NoStop}{Audoly1999}%
\bibitem{Coupeau2002}%
  \BibitemOpen
  \bibfield{author}{%
  \bibinfo {author} {\bibfnamefont{C.}~\bibnamefont{Coupeau}},\ }%
  \bibfield{title}{%
  \enquote{\bibinfo {title} {Atomic force microscopy study of the morphological
  shape of thin film buckling},}\ }%
  \bibfield{journal}{%
  \bibinfo {journal} {Thin Solid Films}\ }%
  \textbf{\bibinfo {volume} {406}},\ \bibinfo {pages} {190} (\bibinfo {year}
  {2002})%
  \bibAnnoteFile{NoStop}{Coupeau2002}%
\bibitem{Kwon2005}%
  \BibitemOpen
  \bibfield{author}{%
  \bibinfo {author} {\bibfnamefont{S.~Joon}\ \bibnamefont{Kwon}}, \bibinfo
  {author} {\bibfnamefont{Jae-Hwan}\ \bibnamefont{Park}},\ and\ \bibinfo
  {author} {\bibfnamefont{Jae-Gwan}\ \bibnamefont{Park}},\ }%
  \bibfield{title}{%
  \enquote{\bibinfo {title} {Wrinkling of a sol-gel-derived thin film},}\ }%
  \bibfield{journal}{%
  \Doi{10.1103/PhysRevE.71.011604}{\bibinfo {journal} {Physical Review E}}\ }%
  \textbf{\bibinfo {volume} {71}},\ \bibinfo {pages} {011604} (\bibinfo {year}
  {2005})%
  \bibAnnoteFile{NoStop}{Kwon2005}%
\bibitem{Edmondson2006}%
  \BibitemOpen
  \bibfield{author}{%
  \bibinfo {author} {\bibfnamefont{S.}~\bibnamefont{Edmondson}}, \bibinfo
  {author} {\bibfnamefont{K.}~\bibnamefont{Frieda}}, \bibinfo {author}
  {\bibfnamefont{J. E.}\ \bibnamefont{Comrie}}, \bibinfo {author}
  {\bibfnamefont{P. R.}\ \bibnamefont{Onck}},\ and\ \bibinfo {author}
  {\bibfnamefont{W. T. S.}\ \bibnamefont{Huck}},\ }%
  \bibfield{title}{%
  \enquote{\bibinfo {title} {Buckling in quasi-2d polymers},}\ }%
  \bibfield{journal}{%
  \Doi{10.1002/adma.200501509}{\bibinfo {journal} {Advanced Materials}}\ }%
  \textbf{\bibinfo {volume} {18}},\ \bibinfo {pages} {724--728} (\bibinfo
  {year} {2006})%
  \bibAnnoteFile{NoStop}{Edmondson2006}%
\bibitem{Jagla2007}%
  \BibitemOpen
  \bibfield{author}{%
  \bibinfo {author} {\bibfnamefont{E.~A.}\ \bibnamefont{Jagla}},\ }%
  \bibfield{title}{%
  \enquote{\bibinfo {title} {Modeling the buckling and delamination of thin
  films},}\ }%
  \bibfield{journal}{%
  \Doi{10.1103/PhysRevB.75.085405}{\bibinfo {journal} {Phys. Rev. B}}\ }%
  \textbf{\bibinfo {volume} {75}},\ \bibinfo {pages} {085405} (\bibinfo {year} {2007})%
  \bibAnnoteFile{NoStop}{Jagla2007}%
\bibitem{Vandeparre2008}%
  \BibitemOpen
  \bibfield{author}{%
  \bibinfo {author} {\bibfnamefont{Hugues}\ \bibnamefont{Vandeparre}}\ and\
  \bibinfo {author} {\bibfnamefont{Pascal}\ \bibnamefont{Damman}},\ }%
  \bibfield{title}{%
  \enquote{\bibinfo {title} {Wrinkling of stimuloresponsive surfaces:
  Mechanical instability coupled to diffusion},}\ }%
  \bibfield{journal}{%
  \Doi{10.1103/PhysRevLett.101.124301}{\bibinfo {journal} {Phys. Rev. Lett.}}\
  }%
  \textbf{\bibinfo {volume} {101}},\ \bibinfo {pages} {124301} (\bibinfo {year} {2008})%
  \bibAnnoteFile{NoStop}{Vandeparre2008}%
\bibitem{Aoyanagi2010}%
  \BibitemOpen
  \bibfield{author}{%
  \bibinfo {author} {\bibfnamefont{Yuko}\ \bibnamefont{Aoyanagi}}, \bibinfo
  {author} {\bibfnamefont{Jeremy}\ \bibnamefont{Hure}}, \bibinfo {author}
  {\bibfnamefont{Jose}\ \bibnamefont{Bico}},\ and\ \bibinfo {author}
  {\bibfnamefont{Benoit}\ \bibnamefont{Roman}},\ }%
  \bibfield{title}{%
  \enquote{\bibinfo {title} {Random blisters on stickers: metrology through
  defects},}\ }%
  \bibfield{journal}{%
  \Doi{10.1039/C0SM00436G}{\bibinfo {journal} {Soft Matter}}\ }%
  \textbf{\bibinfo {volume} {6}},\ \bibinfo {pages} {5720--5728} (\bibinfo
  {year} {2010})%
  \bibAnnoteFile{NoStop}{Aoyanagi2010}%
\bibitem{Annabattula2011}%
  \BibitemOpen
  \bibfield{author}{%
  \bibinfo {author} {\bibfnamefont{R.~K.}\ \bibnamefont{Annabattula}}\ and\
  \bibinfo {author} {\bibfnamefont{P.~R.}\ \bibnamefont{Onck}},\ }%
  \bibfield{title}{%
  \enquote{\bibinfo {title} {Micron-scale pattern formation in prestressed
  polygonal films},}\ }%
  \bibfield{journal}{%
  \bibinfo {journal} {Journal of Applied Physics}\ }%
  \textbf{\bibinfo {volume} {109}},\ \bibinfo {eid} {033517} (\bibinfo {year}
  {2011})%
  \bibAnnoteFile{NoStop}{Annabattula2011}%
\bibitem{Faou2012}%
  \BibitemOpen
  \bibfield{author}{%
  \bibinfo {author} {\bibfnamefont{Jean-Yvon}\ \bibnamefont{Faou}}, \bibinfo
  {author} {\bibfnamefont{Guillaume}\ \bibnamefont{Parry}}, \bibinfo {author}
  {\bibfnamefont{Sergey}\ \bibnamefont{Grachev}},\ and\ \bibinfo {author}
  {\bibfnamefont{Etienne}\ \bibnamefont{Barthel}},\ }%
  \bibfield{title}{%
  \enquote{\bibinfo {title} {How does adhesion induce the formation of
  telephone cord buckles?}.}\ }%
  \bibfield{journal}{%
  \Doi{10.1103/PhysRevLett.108.116102}{\bibinfo {journal} {Phys. Rev. Lett.}}\
  }%
  \textbf{\bibinfo {volume} {108}},\ \bibinfo {pages} {116102} (\bibinfo {year} {2012})%
  \bibAnnoteFile{NoStop}{Faou2012}%
\bibitem{Yu2013}%
  \BibitemOpen
  \bibfield{author}{%
  \bibinfo {author} {\bibfnamefont{Sen-Jiang}\ \bibnamefont{Yu}}, \bibinfo
  {author} {\bibfnamefont{Yuan-Chi}\ \bibnamefont{Shi}}, \bibinfo {author}
  {\bibfnamefont{Miao-Gen}\ \bibnamefont{Chen}}, \bibinfo {author}
  {\bibfnamefont{Ping-Zhan}\ \bibnamefont{Si}}, \bibinfo {author}
  {\bibfnamefont{Yun}\ \bibnamefont{Zhou}}, \bibinfo {author}
  {\bibfnamefont{Xiao-Fei}\ \bibnamefont{Zhang}}, \bibinfo {author}
  {\bibfnamefont{Jun}\ \bibnamefont{Chen}}, \bibinfo {author}
  {\bibfnamefont{Hong}\ \bibnamefont{Zhou}},\ and\ \bibinfo {author}
  {\bibfnamefont{Zhi-Wei}\ \bibnamefont{Jiao}},\ }%
  \bibfield{title}{%
  \enquote{\bibinfo {title} {Coalescence behaviors of telephone cord buckles in
  SiAlN$_x$ films},}\ }%
  \bibfield{journal}{%
  \Doi{http://dx.doi.org/10.1016/j.surfcoat.2013.06.117}{\bibinfo {journal}
  {Surface and Coatings Technology}}\ }%
  \textbf{\bibinfo {volume} {232}},\ \bibinfo {pages} {884 -- 890} (\bibinfo
  {year} {2013})%
  \bibAnnoteFile{NoStop}{Yu2013}%
\bibitem{Tomori2011}%
  \BibitemOpen
  \bibfield{author}{%
  \bibinfo {author} {\bibfnamefont{Hikari}\ \bibnamefont{Tomori}}, \bibinfo
  {author} {\bibfnamefont{Akinobu}\ \bibnamefont{Kanda}}, \bibinfo {author}
  {\bibfnamefont{Hidenori}\ \bibnamefont{Goto}}, \bibinfo {author}
  {\bibfnamefont{Youiti}\ \bibnamefont{Ootuka}}, \bibinfo {author}
  {\bibfnamefont{Kazuhito}\ \bibnamefont{Tsukagoshi}}, \bibinfo {author}
  {\bibfnamefont{Satoshi}\ \bibnamefont{Moriyama}}, \bibinfo {author}
  {\bibfnamefont{Eiichiro}\ \bibnamefont{Watanabe}},\ and\ \bibinfo {author}
  {\bibfnamefont{Daiju}\ \bibnamefont{Tsuya}},\ }%
  \bibfield{title}{%
  \enquote{\bibinfo {title} {Introducing nonuniform strain to graphene using
  dielectric nanopillars},}\ }%
  \bibfield{journal}{%
  \bibinfo {journal} {Applied Physics Express}\ }%
  \textbf{\bibinfo {volume} {4}},\ \bibinfo {pages} {075102} (\bibinfo {year}
  {2011})%
  \bibAnnoteFile{NoStop}{Tomori2011}%
\bibitem{NeekAmal2012}%
  \BibitemOpen
  \bibfield{author}{%
  \bibinfo {author} {\bibfnamefont{M.}~\bibnamefont{Neek-Amal}}, \bibinfo
  {author} {\bibfnamefont{L.}~\bibnamefont{Covaci}},\ and\ \bibinfo {author}
  {\bibfnamefont{F.~M.}\ \bibnamefont{Peeters}},\ }%
  \bibfield{title}{%
  \enquote{\bibinfo {title} {Nanoengineered nonuniform strain in graphene using
  nanopillars},}\ }%
  \bibfield{journal}{%
  \Doi{10.1103/PhysRevB.86.041405}{\bibinfo {journal} {Phys. Rev. B}}\ }%
  \textbf{\bibinfo {volume} {86}},\ \bibinfo {pages} {041405} (\bibinfo {year} {2012})%
  \bibAnnoteFile{NoStop}{NeekAmal2012}%
\bibitem{Yamamoto2012}%
  \BibitemOpen
  \bibfield{author}{%
  \bibinfo {author} {\bibfnamefont{Mahito}\ \bibnamefont{Yamamoto}}, \bibinfo
  {author} {\bibfnamefont{Olivier}\ \bibnamefont{Pierre-Louis}}, \bibinfo
  {author} {\bibfnamefont{Jia}\ \bibnamefont{Huang}}, \bibinfo {author}
  {\bibfnamefont{Michael~S.}\ \bibnamefont{Fuhrer}}, \bibinfo {author}
  {\bibfnamefont{Theodore~L.}\ \bibnamefont{Einstein}},\ and\ \bibinfo {author}
  {\bibfnamefont{William~G.}\ \bibnamefont{Cullen}},\ }%
  \bibfield{title}{%
  \enquote{\bibinfo {title} {{``The Princess and the Pea'' at the Nanoscale:
  Wrinkling and Delamination of Graphene on Nanoparticles}},}\ }%
  \bibfield{journal}{%
  \Doi{10.1103/PhysRevX.2.041018}{\bibinfo {journal} {Phys. Rev. X}}\ }%
  \textbf{\bibinfo {volume} {2}},\ \bibinfo {pages} {041018} (\bibinfo {year} {2012})%
  \bibAnnoteFile{NoStop}{Yamamoto2012}%
\bibitem{Zhu2014}%
  \BibitemOpen
  \bibfield{author}{%
  \bibinfo {author} {\bibfnamefont{Shuze}\ \bibnamefont{Zhu}}\ and\ \bibinfo
  {author} {\bibfnamefont{Teng}\ \bibnamefont{Li}},\ }%
  \bibfield{title}{%
  \enquote{\bibinfo {title} {Wrinkling instability of graphene on
  substrate-supported nanoparticles},}\ }%
  \bibfield{journal}{%
  \Doi{10.1115/1.4026638}{\bibinfo {journal} {Journal of Applied Mechanics}}\
  }%
  \textbf{\bibinfo {volume} {81}},\ \bibinfo {pages} {061008} (\bibinfo {year}
  {2014})%
  \bibAnnoteFile{NoStop}{Zhu2014}%
\bibitem{ViolaKusminskiy2011}%
  \BibitemOpen
  \bibfield{author}{%
  \bibinfo {author} {\bibfnamefont{S.}~\bibnamefont{Viola~Kusminskiy}},
  \bibinfo {author} {\bibfnamefont{D.~K}\ \bibnamefont{Campbell}}, \bibinfo
  {author} {\bibfnamefont{A.~H.}\ \bibnamefont{Castro~Neto}},\ and\ \bibinfo
  {author} {\bibfnamefont{F.}~\bibnamefont{Guinea}},\ }%
  \bibfield{title}{%
  \enquote{\bibinfo {title} {Pinning of a two-dimensional membrane on top of a
  patterned substrate: The case of graphene},}\ }%
  \bibfield{journal}{%
  \Doi{10.1103/PhysRevB.83.165405}{\bibinfo {journal} {Phys. Rev. B}}\ }%
  \textbf{\bibinfo {volume} {83}},\ \bibinfo {pages} {165405} (\bibinfo {year} {2011})%
  \bibAnnoteFile{NoStop}{ViolaKusminskiy2011}%
\bibitem{Kranz1979}%
  \BibitemOpen
  \bibfield{author}{%
  \bibinfo {author} {\bibfnamefont{Robert~L.}\ \bibnamefont{Kranz}},\ }%
  \bibfield{title}{%
  \enquote{\bibinfo {title} {Crack-crack and crack-pore interactions in
  stressed granite},}\ }%
  \bibfield{journal}{%
  \Doi{http://dx.doi.org/10.1016/0148-9062(79)90773-3}{\bibinfo {journal}
  {International Journal of Rock Mechanics and Mining Sciences {\&}
  Geomechanics Abstracts}}\ }%
  \textbf{\bibinfo {volume} {16}},\ \bibinfo {pages} {37 -- 47} (\bibinfo
  {year} {1979})%
  \bibAnnoteFile{NoStop}{Kranz1979}%
\bibitem{Fender2010}%
  \BibitemOpen
  \bibfield{author}{%
  \bibinfo {author} {\bibfnamefont{Melissa~L.}\ \bibnamefont{Fender}}, \bibinfo
  {author} {\bibfnamefont{Fr\'ed\'eric}\ \bibnamefont{Lechenault}},\ and\
  \bibinfo {author} {\bibfnamefont{Karen~E.}\ \bibnamefont{Daniels}},\ }%
  \bibfield{title}{%
  \enquote{\bibinfo {title} {Universal shapes formed by two interacting
  cracks},}\ }%
  \bibfield{journal}{%
  \Doi{10.1103/PhysRevLett.105.125505}{\bibinfo {journal} {Phys. Rev. Lett.}}\
  }%
  \textbf{\bibinfo {volume} {105}},\ \bibinfo {pages} {125505} (\bibinfo {year} {2010})%
  \bibAnnoteFile{NoStop}{Fender2010}%
\bibitem{Fasolino2007}%
  \BibitemOpen
  \bibfield{author}{%
  \bibinfo {author} {\bibfnamefont{A.}~\bibnamefont{Fasolino}}, \bibinfo
  {author} {\bibfnamefont{J.~H.}\ \bibnamefont{Los}},\ and\ \bibinfo {author}
  {\bibfnamefont{M.~I.}\ \bibnamefont{Katsnelson}},\ }%
  \bibfield{title}{%
  \enquote{\bibinfo {title} {Intrinsic ripples in graphene},}\ }%
  \bibfield{journal}{%
  \Doi{10.1038/nmat2011}{\bibinfo {journal} {Nature Materials}}\ }%
  \textbf{\bibinfo {volume} {6}},\ \bibinfo {pages} {858} (\bibinfo {year}
  {2007})%
  \bibAnnoteFile{NoStop}{Fasolino2007}%
\bibitem{Lee2012}%
  \BibitemOpen
  \bibfield{author}{%
  \bibinfo {author} {\bibfnamefont{Jae-Ung}\ \bibnamefont{Lee}}, \bibinfo
  {author} {\bibfnamefont{Duhee}\ \bibnamefont{Yoon}},\ and\ \bibinfo {author}
  {\bibfnamefont{Hyeonsik}\ \bibnamefont{Cheong}},\ }%
  \bibfield{title}{%
  \enquote{\bibinfo {title} {Estimation of young’s modulus of graphene by
  raman spectroscopy},}\ }%
  \bibfield{journal}{%
  \Doi{10.1021/nl301073q}{\bibinfo {journal} {Nano Letters}}\ }%
  \textbf{\bibinfo {volume} {12}},\ \bibinfo {pages} {4444--4448} (\bibinfo
  {year} {2012})%
  \bibAnnoteFile{NoStop}{Lee2012}%
\bibitem{Hartmann2013}%
  \BibitemOpen
  \bibfield{author}{%
  \bibinfo {author} {\bibfnamefont{Markus~A.}\ \bibnamefont{Hartmann}},
  \bibinfo {author} {\bibfnamefont{Melanie}\ \bibnamefont{Todt}}, \bibinfo
  {author} {\bibfnamefont{Franz~G.}\ \bibnamefont{Rammerstorfer}}, \bibinfo
  {author} {\bibfnamefont{Franz~D.}\ \bibnamefont{Fischer}},\ and\ \bibinfo
  {author} {\bibfnamefont{Oskar}\ \bibnamefont{Paris}},\ }%
  \bibfield{title}{%
  \enquote{\bibinfo {title} {Elastic properties of graphene obtained by
  computational mechanical tests},}\ }%
  \bibfield{journal}{%
  \bibinfo {journal} {E. P. L.}\ }%
  \textbf{\bibinfo {volume} {103}},\ \bibinfo {pages} {68004} (\bibinfo {year}
  {2013})%
  \bibAnnoteFile{NoStop}{Hartmann2013}%
\bibitem{Novoselov2004}%
  \BibitemOpen
  \bibfield{author}{%
  \bibinfo {author} {\bibfnamefont{K.~S.}\ \bibnamefont{Novoselov}}, \bibinfo
  {author} {\bibfnamefont{A.~K.}\ \bibnamefont{Geim}}, \bibinfo {author}
  {\bibfnamefont{S.~V.}\ \bibnamefont{Morozov}}, \bibinfo {author}
  {\bibfnamefont{D.}~\bibnamefont{Jiang}}, \bibinfo {author}
  {\bibfnamefont{Y.}~\bibnamefont{Zhang}}, \bibinfo {author}
  {\bibfnamefont{S.~V.}\ \bibnamefont{Dubonos}}, \bibinfo {author}
  {\bibfnamefont{I.~V.}\ \bibnamefont{Grigorieva}},\ and\ \bibinfo {author}
  {\bibfnamefont{A.~A.}\ \bibnamefont{Firsov}},\ }%
  \bibfield{title}{%
  \enquote{\bibinfo {title} {Electric field effect in atomically thin carbon
  films},}\ }%
  \bibfield{journal}{%
  \Doi{10.1126/science.1102896}{\bibinfo {journal} {Science}}\ }%
  \textbf{\bibinfo {volume} {306}},\ \bibinfo {pages} {666--669} (\bibinfo
  {year} {2004})%
  \bibAnnoteFile{NoStop}{Novoselov2004}%
\bibitem{CastroNeto2009}%
  \BibitemOpen
  \bibfield{author}{%
  \bibinfo {author} {\bibfnamefont{A.~H.}\ \bibnamefont{Castro~Neto}}, \bibinfo
  {author} {\bibfnamefont{F.}~\bibnamefont{Guinea}}, \bibinfo {author}
  {\bibfnamefont{N.~M.~R.}\ \bibnamefont{Peres}}, \bibinfo {author}
  {\bibfnamefont{K.~S.}\ \bibnamefont{Novoselov}},\ and\ \bibinfo {author}
  {\bibfnamefont{A.~K.}\ \bibnamefont{Geim}},\ }%
  \bibfield{title}{%
  \enquote{\bibinfo {title} {The electronic properties of graphene},}\ }%
  \bibfield{journal}{%
  \Doi{10.1103/RevModPhys.81.109}{\bibinfo {journal} {Rev. Mod. Phys.}}\ }%
  \textbf{\bibinfo {volume} {81}},\ \bibinfo {pages} {109--162} (\bibinfo {year} {2009})%
  \bibAnnoteFile{NoStop}{CastroNeto2009}%
\bibitem{Novoselov2012}%
  \BibitemOpen
  \bibfield{author}{%
  \bibinfo {author} {\bibfnamefont{K.~S.}\ \bibnamefont{Novoselov}}, \bibinfo
  {author} {\bibfnamefont{Fal'ko~V.}\ \bibnamefont{I.}}, \bibinfo {author}
  {\bibfnamefont{Colombo}\ \bibnamefont{L.}}, \bibinfo {author}
  {\bibfnamefont{Gellert~P.}\ \bibnamefont{R.}}, \bibinfo {author}
  {\bibfnamefont{Schwab~M.}\ \bibnamefont{G.}},\ and\ \bibinfo {author}
  {\bibfnamefont{Kim}\ \bibnamefont{K.}},\ }%
  \bibfield{title}{%
  \enquote{\bibinfo {title} {A roadmap for graphene},}\ }%
  \bibfield{journal}{%
  \Doi{http://dx.doi.org/10.1038/nature11458}{\bibinfo {journal} {Nature}}\ }%
  \textbf{\bibinfo {volume} {490}},\ \bibinfo {pages} {192--200} (\bibinfo {year} {2012})%
  \bibAnnoteFile{NoStop}{Novoselov2012}%
\bibitem{Pereira2009}%
  \BibitemOpen
  \bibfield{author}{%
  \bibinfo {author} {\bibfnamefont{Vitor~M.}\ \bibnamefont{Pereira}}\ and\
  \bibinfo {author} {\bibfnamefont{A.~H.}\ \bibnamefont{Castro~Neto}},\ }%
  \bibfield{title}{%
  \enquote{\bibinfo {title} {Strain engineering of graphene's electronic
  structure},}\ }%
  \bibfield{journal}{%
  \Doi{10.1103/PhysRevLett.103.046801}{\bibinfo {journal} {Phys. Rev. Lett.}}\
  }%
  \textbf{\bibinfo {volume} {103}},\ \bibinfo {pages} {046801} (\bibinfo {year} {2009})%
  \bibAnnoteFile{NoStop}{Pereira2009}%
\bibitem{Wei2011}%
  \BibitemOpen
  \bibfield{author}{%
  \bibinfo {author} {\bibfnamefont{Ning}\ \bibnamefont{Wei}}, \bibinfo {author}
  {\bibfnamefont{Lanqing}\ \bibnamefont{Xu}}, \bibinfo {author}
  {\bibfnamefont{Hui-Qiong}\ \bibnamefont{Wang}},\ and\ \bibinfo {author}
  {\bibfnamefont{Jin-Cheng}\ \bibnamefont{Zheng}},\ }%
  \bibfield{title}{%
  \enquote{\bibinfo {title} {Strain engineering of thermal conductivity in
  graphene sheets and nanoribbons: a demonstration of magic flexibility},}\ }%
  \bibfield{journal}{%
  \bibinfo {journal} {Nanotechnology}\ }%
  \textbf{\bibinfo {volume} {22}},\ \bibinfo {pages} {105705} (\bibinfo {year}
  {2011})%
  \bibAnnoteFile{NoStop}{Wei2011}%
\bibitem{Ni2014}%
  \BibitemOpen
  \bibfield{author}{%
  \bibinfo {author} {\bibfnamefont{Guang-Xin}\ \bibnamefont{Ni}}, \bibinfo
  {author} {\bibfnamefont{Hong-Zhi}\ \bibnamefont{Yang}}, \bibinfo {author}
  {\bibfnamefont{Wei}\ \bibnamefont{Ji}}, \bibinfo {author}
  {\bibfnamefont{Seung-Jae}\ \bibnamefont{Baeck}}, \bibinfo {author}
  {\bibfnamefont{Chee-Tat}\ \bibnamefont{Toh}}, \bibinfo {author}
  {\bibfnamefont{Jong-Hyun}\ \bibnamefont{Ahn}}, \bibinfo {author}
  {\bibfnamefont{Vitor~M.}\ \bibnamefont{Pereira}},\ and\ \bibinfo {author}
  {\bibfnamefont{Barbaros}\ \bibnamefont{{\"Ozyilmaz}}},\ }%
  \bibfield{title}{%
  \enquote{\bibinfo {title} {Tuning optical conductivity of large-scale cvd
  graphene by strain engineering},}\ }%
  \bibfield{journal}{%
  \Doi{10.1002/adma.201304156}{\bibinfo {journal} {Advanced Materials}}\ }%
  \textbf{\bibinfo {volume} {26}},\ \bibinfo {pages} {1081--1086} (\bibinfo
  {year} {2014})%
  \bibAnnoteFile{NoStop}{Ni2014}%
\bibitem{Sutter2013}%
  \BibitemOpen
  \bibfield{author}{%
  \bibinfo {author} {\bibfnamefont{Peter}\ \bibnamefont{Sutter}}, \bibinfo
  {author} {\bibfnamefont{Peter}\ \bibnamefont{Albrecht}}, \bibinfo {author}
  {\bibfnamefont{Xiao}\ \bibnamefont{Tong}},\ and\ \bibinfo {author}
  {\bibfnamefont{Eli}\ \bibnamefont{Sutter}},\ }%
  \bibfield{title}{%
  \enquote{\bibinfo {title} {Mechanical decoupling of graphene from Ru(0001) by
  interfacial reaction with oxygen},}\ }%
  \bibfield{journal}{%
  \Doi{10.1021/jp400838j}{\bibinfo {journal} {The Journal of Physical Chemistry
  C}}\ }%
  \textbf{\bibinfo {volume} {117}},\ \bibinfo {pages} {6320--6324} (\bibinfo
  {year} {2013})%
  \bibAnnoteFile{NoStop}{Sutter2013}%
\bibitem{Matuda1981}%
  \BibitemOpen
  \bibfield{author}{%
  \bibinfo {author} {\bibfnamefont{N.}~\bibnamefont{Matuda}}, \bibinfo {author}
  {\bibfnamefont{S.}~\bibnamefont{Baba}},\ and\ \bibinfo {author}
  {\bibfnamefont{A.}~\bibnamefont{Kinbara}},\ }%
  \bibfield{title}{%
  \enquote{\bibinfo {title} {Internal stress, young's modulus and adhesion
  energy of carbon films on glass substrates},}\ }%
  \bibfield{journal}{%
  \bibinfo {journal} {Thin Solid Films}\ }%
  \textbf{\bibinfo {volume} {81}},\ \bibinfo {pages} {301} (\bibinfo {year}
  {1981})%
  \bibAnnoteFile{NoStop}{Matuda1981}%
\bibitem{Take2007}%
  \BibitemOpen
  \bibfield{author}{%
  \bibinfo {author} {\bibfnamefont{W.~A.}\ \bibnamefont{Take}}, \bibinfo
  {author} {\bibfnamefont{M.~J.}\ \bibnamefont{Chappel}}, \bibinfo {author}
  {\bibfnamefont{R.~W.~I.}\ \bibnamefont{Brachman}},\ and\ \bibinfo {author}
  {\bibfnamefont{R.~K.}\ \bibnamefont{Rowe}},\ }%
  \bibfield{title}{%
  \enquote{\bibinfo {title} {Quantifying geomembrane wrinkles using aerial
  photography and digital image processing},}\ }%
  \bibfield{journal}{%
  \Doi{10.1680/gein.2007.14.4.219}{\bibinfo {journal} {Geosynthetics
  International}}\ }%
  \textbf{\bibinfo {volume} {14}},\ \bibinfo {pages} {219} (\bibinfo {year}
  {2007})%
  \bibAnnoteFile{NoStop}{Take2007}%
\bibitem{Poot2008}%
  \BibitemOpen
  \bibfield{author}{%
  \bibinfo {author} {\bibfnamefont{M.}~\bibnamefont{Poot}}\ and\ \bibinfo
  {author} {\bibfnamefont{H.~S.~J.}\ \bibnamefont{van~der Zant}},\ }%
  \bibfield{title}{%
  \enquote{\bibinfo {title} {Nanomechanical properties of few-layer graphene
  membranes},}\ }%
  \bibfield{journal}{%
  \bibinfo {journal} {Applied Physics Letters}\ }%
  \textbf{\bibinfo {volume} {92}},\ \bibinfo {eid} {063111} (\bibinfo {year}
  {2008})%
  \bibAnnoteFile{NoStop}{Poot2008}%
\bibitem{Plimpton1995}%
  \BibitemOpen
  \bibfield{author}{%
  \bibinfo {author} {\bibfnamefont{S.}~\bibnamefont{Plimpton}},\ }%
  \bibfield{title}{%
  \enquote{\bibinfo {title} {Fast parallel algorithms for short-range molecular
  dynamics},}\ }%
  \bibfield{journal}{%
  \bibinfo {journal} {J. Comp. Phys.}\ }%
  \textbf{\bibinfo {volume} {117}},\ \bibinfo {pages} {1--19} (\bibinfo {year}
  {1995})%
  \bibAnnoteFile{NoStop}{Plimpton1995}%
\bibitem{SuppMat}%
  \BibitemOpen
  \ \bibinfo {note} {{See Supplemental Material at URL for examples of wrinkle
  propagation, deposition of graphene on four nanoparticles, and deposition of
  atomistic graphene on a substrate with which it interacts {\it via} a radial
  Lennard-Jones potential.}}%
  \bibAnnoteFile{Stop}{SuppMat}%
\bibitem{LandauLifshitz}%
  \BibitemOpen
  \bibfield{author}{%
  \bibinfo {author} {\bibfnamefont{L.D.}\ \bibnamefont{Landau}}\ and\ \bibinfo
  {author} {\bibfnamefont{E.M.}\ \bibnamefont{Lifshitz}},\ }%
  \emph{\bibinfo {title} {Theory of Elasticity}}\ (\bibinfo {publisher}
  {Butterworth-Heinemann},\ \bibinfo {address} {Oxford},\ \bibinfo {year}
  {1986})%
  \bibAnnoteFile{NoStop}{LandauLifshitz}%
\bibitem{Zong2010}%
  \BibitemOpen
  \bibfield{author}{%
  \bibinfo {author} {\bibfnamefont{Zong}\ \bibnamefont{Zong}}, \bibinfo
  {author} {\bibfnamefont{Chia-Ling}\ \bibnamefont{Chen}}, \bibinfo {author}
  {\bibfnamefont{Mehmet~R.}\ \bibnamefont{Dokmeci}},\ and\ \bibinfo {author}
  {\bibfnamefont{Kai-tak}\ \bibnamefont{Wan}},\ }%
  \bibfield{title}{%
  \enquote{\bibinfo {title} {Direct measurement of graphene adhesion on silicon
  surface by intercalation of nanoparticles},}\ }%
  \bibfield{journal}{%
  \bibinfo {journal} {Journal of Applied Physics}\ }%
  \textbf{\bibinfo {volume} {107}},\ \bibinfo {eid} {026104} (\bibinfo {year}
  {2010})%
  \bibAnnoteFile{NoStop}{Zong2010}%
\bibitem{Cortet2008}%
  \BibitemOpen
  \bibfield{author}{%
  \bibinfo {author} {\bibfnamefont{Pierre-Philippe}\ \bibnamefont{Cortet}},
  \bibinfo {author} {\bibfnamefont{Guillaume}\ \bibnamefont{Huillard}},
  \bibinfo {author} {\bibfnamefont{Lo\"ic}\ \bibnamefont{Vanel}},\ and\ \bibinfo
  {author} {\bibfnamefont{Sergio}\ \bibnamefont{Ciliberto}},\ }%
  \bibfield{title}{%
  \enquote{\bibinfo {title} {Attractive and repulsive cracks in a heterogeneous
  material},}\ }%
  \bibfield{journal}{%
  \bibinfo {journal} {Journal of Statistical Mechanics: Theory and Experiment}\
  }%
  \textbf{\bibinfo {volume} {2008}},\ \bibinfo {pages} {P10022} (\bibinfo
  {year} {2008})%
  \bibAnnoteFile{NoStop}{Cortet2008}%
\bibitem{deArcangelis1985}%
  \BibitemOpen
  \bibfield{author}{%
  \bibinfo {author} {\bibnamefont{{de Arcangelis, L.}}}, \bibinfo {author}
  {\bibnamefont{{Redner, S.}}},\ and\ \bibinfo {author}
  {\bibnamefont{{Herrmann, H.J.}}},\ }%
  \bibfield{title}{%
  \enquote{\bibinfo {title} {A random fuse model for breaking processes},}\ }%
  \bibfield{journal}{%
  \Doi{10.1051/jphyslet:019850046013058500}{\bibinfo {journal} {J. Physique
  Lett.}}\ }%
  \textbf{\bibinfo {volume} {46}},\ \bibinfo {pages} {585} (\bibinfo {year}
  {1985})%
  \bibAnnoteFile{NoStop}{deArcangelis1985}%
\bibitem{Duxbury1987}%
  \BibitemOpen
  \bibfield{author}{%
  \bibinfo {author} {\bibfnamefont{P.~M.}\ \bibnamefont{Duxbury}}, \bibinfo
  {author} {\bibfnamefont{P.~L.}\ \bibnamefont{Leath}},\ and\ \bibinfo {author}
  {\bibfnamefont{Paul~D.}\ \bibnamefont{Beale}},\ }%
  \bibfield{title}{%
  \enquote{\bibinfo {title} {Breakdown properties of quenched random systems:
  The random-fuse network},}\ }%
  \bibfield{journal}{%
  \Doi{10.1103/PhysRevB.36.367}{\bibinfo {journal} {Phys. Rev. B}}\ }%
  \textbf{\bibinfo {volume} {36}},\ \bibinfo {pages} {367--380} (\bibinfo {year} {1987})%
  \bibAnnoteFile{NoStop}{Duxbury1987}%
\bibitem{PierreLouis2008}%
  \BibitemOpen
  \bibfield{author}{%
  \bibinfo {author} {\bibfnamefont{O.}~\bibnamefont{Pierre-Louis}},\ }%
  \bibfield{title}{%
  \enquote{\bibinfo {title} {Adhesion of membranes and filaments on rippled
  surfaces},}\ }%
  \bibfield{journal}{%
  \Doi{10.1103/PhysRevE.78.021603}{\bibinfo {journal} {Phys. Rev. E}}\ }%
  \textbf{\bibinfo {volume} {78}},\ \bibinfo {pages} {021603} (\bibinfo {year} {2008})%
  \bibAnnoteFile{NoStop}{PierreLouis2008}%
\bibitem{Cullen2010}%
  \BibitemOpen
  \bibfield{author}{%
  \bibinfo {author} {\bibfnamefont{W.~G.}\ \bibnamefont{Cullen}}, \bibinfo
  {author} {\bibfnamefont{M.}~\bibnamefont{Yamamoto}}, \bibinfo {author}
  {\bibfnamefont{K.~M.}\ \bibnamefont{Burson}}, \bibinfo {author}
  {\bibfnamefont{J.~H.}\ \bibnamefont{Chen}}, \bibinfo {author}
  {\bibfnamefont{C.}~\bibnamefont{Jang}}, \bibinfo {author}
  {\bibfnamefont{L.}~\bibnamefont{Li}}, \bibinfo {author}
  {\bibfnamefont{M.~S.}\ \bibnamefont{Fuhrer}},\ and\ \bibinfo {author}
  {\bibfnamefont{E.~D.}\ \bibnamefont{Williams}},\ }%
  \bibfield{title}{%
  \enquote{\bibinfo {title} {{High-Fidelity Conformation of Graphene to
  SiO$_{2}$ Topographic Features}},}\ }%
  \bibfield{journal}{%
  \Doi{10.1103/PhysRevLett.105.215504}{\bibinfo {journal} {Phys. Rev. Lett.}}\
  }%
  \textbf{\bibinfo {volume} {105}},\ \bibinfo {pages} {215504} (\bibinfo {year} {2010})%
  \bibAnnoteFile{NoStop}{Cullen2010}%
\bibitem{Li2010}%
  \BibitemOpen
  \bibfield{author}{%
  \bibinfo {author} {\bibfnamefont{Teng}\ \bibnamefont{Li}}\ and\ \bibinfo
  {author} {\bibfnamefont{Zhao}\ \bibnamefont{Zhang}},\ }%
  \bibfield{title}{%
  \enquote{\bibinfo {title} {Snap-through instability of graphene on
  substrates},}\ }%
  \bibfield{journal}{%
  \Doi{10.1007/s11671-009-9460-1}{\bibinfo {journal} {Nanoscale Res. Lett.}}\
  }%
  \textbf{\bibinfo {volume} {5}},\ \bibinfo {pages} {169} (\bibinfo {year}
  {2010})%
  \bibAnnoteFile{NoStop}{Li2010}%
\bibitem{NeekAmal2010}%
  \BibitemOpen
  \bibfield{author}{%
  \bibinfo {author} {\bibfnamefont{M.}~\bibnamefont{Neek-Amal}}\ and\ \bibinfo
  {author} {\bibfnamefont{F.~M.}\ \bibnamefont{Peeters}},\ }%
  \bibfield{title}{%
  \enquote{\bibinfo {title} {Nanoindentation of a circular sheet of bilayer
  graphene},}\ }%
  \bibfield{journal}{%
  \Doi{10.1103/PhysRevB.81.235421}{\bibinfo {journal} {Phys. Rev. B}}\ }%
  \textbf{\bibinfo {volume} {81}},\ \bibinfo {pages} {235421} (\bibinfo {year} {2010})%
  \bibAnnoteFile{NoStop}{NeekAmal2010}%
\bibitem{Wagner2012}%
  \BibitemOpen
  \bibfield{author}{%
  \bibinfo {author} {\bibfnamefont{Till J.~W.}\ \bibnamefont{Wagner}}\ and\
  \bibinfo {author} {\bibfnamefont{Dominic}\ \bibnamefont{Vella}},\ }%
  \bibfield{title}{%
  \enquote{\bibinfo {title} {The sensitivity of graphene ``snap-through'' to
  substrate geometry},}\ }%
  \bibfield{journal}{%
  \bibinfo {journal} {Applied Physics Letters}\ }%
  \textbf{\bibinfo {volume} {100}},\ \bibinfo {eid} {233111} (\bibinfo {year}
  {2012})%
  \bibAnnoteFile{NoStop}{Wagner2012}%
\bibitem{Tersoff1988}%
  \BibitemOpen
  \bibfield{author}{%
  \bibinfo {author} {\bibfnamefont{J.}~\bibnamefont{Tersoff}},\ }%
  \bibfield{title}{%
  \enquote{\bibinfo {title} {New empirical approach for the structure and
  energy of covalent systems},}\ }%
  \bibfield{journal}{%
  \Doi{10.1103/PhysRevB.37.6991}{\bibinfo {journal} {Phys. Rev. B}}\ }%
  \textbf{\bibinfo {volume} {37}},\ \bibinfo {pages} {6991--7000} (\bibinfo {year} {1988})%
  \bibAnnoteFile{NoStop}{Tersoff1988}%
\bibitem{Nukala2003}%
  \BibitemOpen
  \bibfield{author}{%
  \bibinfo {author} {\bibfnamefont{Phani Kumar V~V}\ \bibnamefont{Nukala}}\
  and\ \bibinfo {author} {\bibfnamefont{Srdan}\ \bibnamefont{Simunovic}},\ }%
  \bibfield{title}{%
  \enquote{\bibinfo {title} {An efficient algorithm for simulating fracture
  using large fuse networks},}\ }%
  \bibfield{journal}{%
  \bibinfo {journal} {Journal of Physics A: Mathematical and General}\ }%
  \textbf{\bibinfo {volume} {36}},\ \bibinfo {pages} {11403} (\bibinfo {year}
  {2003})%
  \bibAnnoteFile{NoStop}{Nukala2003}%
\bibitem{Koenig2011}%
  \BibitemOpen
  \bibfield{author}{%
  \bibinfo {author} {\bibfnamefont{Steven~P.}\ \bibnamefont{Koenig}}, \bibinfo
  {author} {\bibfnamefont{Narasimha~G.}\ \bibnamefont{Boddeti}}, \bibinfo
  {author} {\bibfnamefont{Martin~L.}\ \bibnamefont{Dunn}},\ and\ \bibinfo
  {author} {\bibfnamefont{J.~Scott}\ \bibnamefont{Bunch}},\ }%
  \bibfield{title}{%
  \enquote{\bibinfo {title} {Ultrastrong adhesion of graphene membranes},}\ }%
  \bibfield{journal}{%
  \bibinfo {journal} {Nature Nanotechnology}\ }%
  \textbf{\bibinfo {volume} {6}},\ \bibinfo {pages} {543} (\bibinfo {year}
  {2011})%
  \bibAnnoteFile{NoStop}{Koenig2011}%
\bibitem{Stuart2000}%
  \BibitemOpen
  \bibfield{author}{%
  \bibinfo {author} {\bibfnamefont{Steven~J.}\ \bibnamefont{Stuart}}, \bibinfo
  {author} {\bibfnamefont{Alan~B.}\ \bibnamefont{Tutein}},\ and\ \bibinfo
  {author} {\bibfnamefont{Judith~A.}\ \bibnamefont{Harrison}},\ }%
  \bibfield{title}{%
  \enquote{\bibinfo {title} {A reactive potential for hydrocarbons with
  intermolecular interactions},}\ }%
  \bibfield{journal}{%
  \Doi{http://dx.doi.org/10.1063/1.481208}{\bibinfo {journal} {The Journal of
  Chemical Physics}}\ }%
  \textbf{\bibinfo {volume} {112}},\ \bibinfo {pages} {6472--6486} (\bibinfo
  {year} {2000})%
  \bibAnnoteFile{NoStop}{Stuart2000}%
\bibitem{brenner2002second}%
  \BibitemOpen
  \bibfield{author}{%
  \bibinfo {author} {\bibfnamefont{Donald~W}\ \bibnamefont{Brenner}}, \bibinfo
  {author} {\bibfnamefont{Olga~A}\ \bibnamefont{Shenderova}}, \bibinfo {author}
  {\bibfnamefont{Judith~A}\ \bibnamefont{Harrison}}, \bibinfo {author}
  {\bibfnamefont{Steven~J}\ \bibnamefont{Stuart}}, \bibinfo {author}
  {\bibfnamefont{Boris}\ \bibnamefont{Ni}},\ and\ \bibinfo {author}
  {\bibfnamefont{Susan~B}\ \bibnamefont{Sinnott}},\ }%
  \bibfield{title}{%
  \enquote{\bibinfo {title} {{A second-generation reactive empirical bond order
  (REBO) potential energy expression for hydrocarbons}},}\ }%
  \bibfield{journal}{%
  \bibinfo {journal} {Journal of Physics: Condensed Matter}\ }%
  \textbf{\bibinfo {volume} {14}},\ \bibinfo {pages} {783} (\bibinfo {year}
  {2002})%
  \bibAnnoteFile{NoStop}{brenner2002second}%
\bibitem{Zhao2009}%
  \BibitemOpen
  \bibfield{author}{%
  \bibinfo {author} {\bibfnamefont{H.}~\bibnamefont{Zhao}}, \bibinfo {author}
  {\bibfnamefont{K.}~\bibnamefont{Min}},\ and\ \bibinfo {author}
  {\bibfnamefont{N.~R.}\ \bibnamefont{Aluru}},\ }%
  \bibfield{title}{%
  \enquote{\bibinfo {title} {Size and chirality dependent elastic properties of
  graphene nanoribbons under uniaxial tension},}\ }%
  \bibfield{journal}{%
  \Doi{10.1021/nl901448z}{\bibinfo {journal} {Nano Letters}}\ }%
  \textbf{\bibinfo {volume} {9}},\ \bibinfo {pages} {3012--3015} (\bibinfo
  {year} {2009})%
  \bibAnnoteFile{NoStop}{Zhao2009}%
\bibitem{Berendsen1984}%
  \BibitemOpen
  \bibfield{author}{%
  \bibinfo {author} {\bibfnamefont{H.~J.~C.}\ \bibnamefont{Berendsen}},
  \bibinfo {author} {\bibfnamefont{J.~P.~M.}\ \bibnamefont{Postma}}, \bibinfo
  {author} {\bibfnamefont{W.~F.}\ \bibnamefont{van Gunsteren}}, \bibinfo
  {author} {\bibfnamefont{A.}~\bibnamefont{DiNola}},\ and\ \bibinfo {author}
  {\bibfnamefont{J.~R.}\ \bibnamefont{Haak}},\ }%
  \bibfield{title}{%
  \enquote{\bibinfo {title} {Molecular dynamics with coupling to an external
  bath},}\ }%
  \bibfield{journal}{%
  \Doi{http://dx.doi.org/10.1063/1.448118}{\bibinfo {journal} {The Journal of
  Chemical Physics}}\ }%
  \textbf{\bibinfo {volume} {81}},\ \bibinfo {pages} {3684--3690} (\bibinfo
  {year} {1984})%
  \bibAnnoteFile{NoStop}{Berendsen1984}%
\end{thebibliography}
\end{document}